\documentclass[useAMS,usenatbib]{mn2e}
\usepackage{psfig}  
\usepackage{amstext,comment}
\usepackage{amssymb}
\usepackage{amsmath, float}
\usepackage{graphicx}
\usepackage{txfonts}
\usepackage{amsmath,mathtools,revsymb,amssymb}
\usepackage{booktabs,multirow,calc,xspace,rotating}
\usepackage{CJK}

\newcommand{\begit}{\begin{itemize}}
\newcommand{\enit}{\end{itemize}}
\newcommand{\begen}{\begin{enumerate}}
\newcommand{\enen}{\end{enumerate}}
\newcommand{\pp}{\partial}    
 
\newcommand{\beq}{\begin{equation}}
\newcommand{\eeq}{\end{equation}}
\newcommand{\beqa}{\begin{eqnarray}} 
\newcommand{\eeqa}{\end{eqnarray}}

\newcommand       \be           {\begin{equation}}
\newcommand       \ee           {\end{equation}}
\newcommand       \bea          {\begin{eqnarray}}
\newcommand       \eea          {\end{eqnarray}}
\newcommand       \kms		{\,{\rm km \,\, s}^{-1}}

\newcommand       \mspy 	{\,{\rm M_\odot \, yr^{-1}}}

\newcommand       \kpc		{\,{\rm kpc }}

\newcommand 	     \Veff		{{\rm V_{g,eff}}}
\newcommand 	     \Veffsq		{{\rm V^2_{g,eff}}}

\begin{document}
\begin{CJK*}{UTF8}{gbsn}
\title[Cosmic-ray Driven Galactic Winds I: Diffusion]{The Physics of Galactic Winds Driven by Cosmic Rays I:  Diffusion}

\author[E. Quataert, T. A. Thompson, Y.F. Jiang]{Eliot Quataert$^{1}$, Todd A. Thompson$^{2,3}$, and Yan-Fei Jiang(姜燕飞)$^{4}$  \\
 $^{1}$Department of Astrophysical Sciences, Princeton University, Princeton, NJ 08544, USA\\ 
$^{2}$ Department of Astronomy, The Ohio State University, 140 West 18th Avenue, Columbus, OH 43210, USA \\
$^{3}$ Center for Cosmology and Astro-Particle Physics (CCAPP), The Ohio State University, 191 West Woodruff Ave., Columbus, OH 43210, USA \\
$^{4}$ Center for Computational Astrophysics, Flatiron Institute, 162 Fifth Avenue, New York, NY, 10010, USA}

\maketitle

\begin{abstract}
The physics of Cosmic ray (CR) transport remains a key uncertainty in assessing whether CRs can produce galaxy-scale outflows consistent with observations.  In this paper, we elucidate the physics of CR-driven galactic winds for CR transport dominated by diffusion. A companion paper considers CR streaming. We use analytic estimates validated by time-dependent spherically-symmetric simulations to derive expressions for the mass-loss rate, momentum flux, and speed of CR-driven galactic winds, suitable for cosmological-scale or semi-analytic models of galaxy formation.  For CR diffusion coefficients $\kappa \gtrsim r_0 c_i$ where $r_0$ is the base radius of the wind and $c_i$ is the isothermal gas sound speed, the asymptotic wind energy flux is comparable to that supplied to CRs, and the outflow rapidly accelerates to supersonic speeds.  By contrast, for $\kappa \lesssim r_0 c_i$, CR-driven winds accelerate more slowly and lose most of their energy to gravity, a CR analogue of photon-tired stellar winds.  Given CR diffusion coefficients estimated using Fermi gamma-ray observations of pion decay, we predict mass-loss rates in CR-driven galactic winds of order the star formation rate for dwarf and disc galaxies.  The dwarf galaxy mass-loss rates are small compared to the mass-loadings needed to reconcile the stellar and dark matter halo mass functions.  For nuclear starbursts (e.g., M82, Arp 220), CR diffusion and pion losses suppress the CR pressure in the galaxy and the strength of CR-driven winds.  We discuss the implications of our results for interpreting observations of galactic winds and for the role of CRs in galaxy formation.
\end{abstract}

\voffset=-2cm

\section{Introduction}
\label{section:introduction}

Galactic winds play a key role in galaxy evolution, shaping the galaxy stellar mass function, the mass-metallicity relation, and affecting the morphology of galaxies over cosmic time (e.g., \citealt{Somerville2015}). Despite their importance, many puzzles persist. These include the acceleration mechanism for the cool atomic and molecular gas seen in emission and absorption from rapidly star-forming galaxies across the universe (e.g., \citealt{Veilleux2020}), the very large mass fluxes and low star formation efficiencies inferred for dwarf galaxies, and the apparent need for massive outflows from normal star-forming galaxies in the local universe in order to match some models of Galactic chemical evolution (e.g., \citealt{Andrews2017}).

Among the physical mechanisms suggested for driving large amounts of cool gas from star-forming galaxies, cosmic rays (CRs) are of particular interest because their pressure is dynamically important with respect to gravity in the Milky Way disc \citep{Boulares1990}. In this picture, CRs are injected into the disc of the galaxy by supernovae and stellar processes, and scatter off of magnetic fluctuations in the ISM, slowly diffusing (or streaming) away from the disc. This process sets up a pressure gradient that can in principle accelerate gas away from the host galaxy.

Many treatments of CR driven winds exist in the literature, starting with the foundational work of \cite{Ipavich1975} who computed time-steady solutions for CR driven winds assuming streaming transport at the Alfv\'en velocity.  More detailed models including CR streaming and hydromagnetic wave pressure were computed by \cite{Breitschwerdt1991}, deriving a mass-loss rate of order M$_\odot$ yr$^{-1}$ from the Galaxy. \cite{Everett2008} further developed a hybrid thermal and CR driven wind model for the Galaxy.   More generally, on the basis of momentum conservation, \cite{Socrates2008} argued that CRs could drive significant mass fluxes from star-forming galaxies.   

In parallel to these more analytic and steady-state treatments, there is a large and growing body of work on multi-dimensional simulations of galaxy formation with CRs, which suggest that they can play a number of important roles: e.g., driving cold-gas outflows from galaxies (e.g., \citealt{Booth2013}), modifying the thermal and ionization state of the CGM (e.g., \citealt{Ji2020}), and heating gas in galaxy groups and clusters, suppressing cooling flows (e.g., \citealt{Ruszkowski2017}).   These numerical models vary in their treatment of  CR transport, including isotropic diffusion (e.g, \citealt{Uhlig2012,Simpson2016}), anisotropic diffusion along magnetic fields (e.g., \citealt{Pakmor2016,Chan2019,Hopkins2020a}), and/or CR streaming along magnetic fields (e.g., \citealt{Ruszkowski2017,Chan2019,Hopkins2020a}).  

As suggested by the diversity of approaches in the literature, one of the significant uncertainties in modeling the properties of CRs in galaxy formation is that the physical mechanism(s) coupling CRs to the thermal plasma are not fully understood (e.g., \citealt{Amato2017}). Small-scale fluctuations in the magnetic field scatter CRs, setting their mean free path. On scales larger than this mean free path, the CRs can be approximated as a fluid \citep{Skilling1971}.   However, it is not clear whether the fluctuations that scatter CRs are self-excited by the CRs themselves (e.g., the streaming instability; \citealt{Kulsrud1969})  or produced by background turbulent fluctuations in the interstellar/circumgalactic-medium (ISM/CGM) cascading to small-scales where they couple to the CRs (e.g., \citealt{Yan2002}).  Nor is it clear whether the dominant scattering mechanism depends on the thermodynamic phase of the ISM/CGM, the local magnetic field strength, or other physical properties. The distinction between self-excited versus background turbulence as the source of scattering that sets the CR mean free path is at the core of the streaming/diffusion dichotomy that dominates CR transport modeling.

\cite{Wiener2017} explored the difference between galactic winds driven with CR diffusion and CR streaming in simulations, finding that diffusive CR transport results in much larger overall mass-loss rates than in CR streaming models.  \cite{Chan2019} and \cite{Hopkins2020a} found similar results and further argued that models with CR transport dominated by diffusion were required for CRs to efficiently escape Milky-way, M31, and Magellanic Cloud-like galaxies, as is required by gamma-ray observations of pion decay produced by CRs interacting with the ISM \citep{Fermi2010,Fermi2010b,Fermi2012}. These results highlight that a better understanding of CR microphysics is critical for understanding the role of CRs in galaxy formation.  

Together with the numerical and analytic treatments of CR-driven winds in the literature, there is also a body of phenomenological work interpreting the non-thermal radio and gamma-ray emission from star-forming galaxies.  This directly informs our understanding of the underlying CR population and their role in driving galactic winds. The far-infrared-radio correlation and the gamma-ray emission from star-forming galaxies, including those with strong winds like M82, NGC 253, and Arp 220 can be used to constrain the average CR injection rate per unit star formation, the gas density seen by CRs as they propagate, the magnetic field strength, and the CR escape timescale (e.g., \citealt{Pavlidou2001, Torres2004, Lacki2010,Lacki2013,Yoast-Hull2013,Yoast-Hull2014,Buckman2020,Crocker2021a,Crocker2020b}). These works thus provide benchmarks for understanding how the inner or ``base" boundary conditions for a putative CR-driven wind vary as a function of galaxy properties.

This paper is the first in a series that aims to elucidate the physics of galactic winds driven by CRs using a combination of analytic calculations and idealized numerical simulations. In this paper we focus on the case of CR diffusion. A companion paper discusses the case of CR streaming, which turns out to be much physically richer than the diffusion limit considered here.  In \S \ref{section:analytic} we present analytic estimates of the mass-loss rate and terminal velocity of galactic winds driven by CRs in the diffusion approximation.  In \S \ref{sec:numerics} we validate these estimates with time-dependent numerical simulations, which further show that the winds reach a laminar steady state.   \S \ref{sec:pion} calibrates CR diffusion coefficients and CR pressures in galaxies using Fermi gamma-ray observations of pion decay, and calculates the resulting implications for CR-driven galactic winds.  We summarize and discuss the implications of our results in \S \ref{sec:disc}.   Appendix \ref{sec:appendixA} presents a linear stability analysis and shows that although sound waves are formally unstable in CR-driven winds with diffusion \citep{Drury1986}, the growth rates are too slow for the instability to be dynamically important in almost all cases (the exception is extremely low gas sound speeds).  We also derive the linear WKB dispersion relation for the two-moment cosmic-ray transport model simulated in \S \ref{sec:numerics} and show that entropy and sound waves are linearly stable in the WKB limit for CR transport by diffusion  in the two-moment CR transport model.

\section{Analytic Approximations for Cosmic Ray Driven Winds with Diffusion}
\label{section:analytic}

In this section we focus on analytic approximations to the spherical steady state galactic wind problem in the presence of cosmic rays.  Following these analytical approximations, in Section \ref{sec:numerics} we treat the more complete numerical problem and solve the time-dependent spherically symmetric CR-driven wind problem using the numerical scheme of \cite{Jiang2018}. We show that the analytic approximations in this section do an excellent job of reproducing the more complete numerical results.

The general equation of motion, including gas pressure $p$, CR pressure $p_{c}$, a general gravitational potential $\Phi$, and magnetic forces in the  limit of magnetohydrodynamics can be written as
\beq
\frac{\pp {\bf v}}{\pp t}+{\bf v}\cdot\nabla {\bf v}=-\frac{1}{\rho}\nabla p-\frac{1}{\rho}\nabla p_{c}-\nabla\Phi+\frac{1}{4\pi \rho}(\nabla\times{\bf B})\times{\bf B}.
\label{motion}
\eeq
The energy equation for the CRs  in the absence of CR sources and pionic losses, and including diffusion and streaming at the Alfv\'en velocity down the CR pressure gradient, i.e., ${\bf v_s} = -{\bf v_A}|\nabla p_c|/\nabla p_c$ with ${\bf v_A}={\bf B}/(4\pi\rho)^{1/2}$, can be written as
\beq
\frac{\pp E_c}{\pp t}+\nabla\cdot {\bf F}_c=\left({\bf v}+{\bf v_{\rm s}}\right)\cdot \nabla p_{c},
\label{cr_energy}
\eeq
where the ``equilibrium" CR flux\footnote{The two-moment CR model we solve numerically in \S \ref{sec:numerics} evolves ${\bf F_c}$ as an independent variable and the flux reduces to equation \ref{equilibrium_flux} only when time variations are sufficiently slow (see eq. \ref{eq:CR2mom}).} is 
\be
{\bf F}_c=4 p_c \left({\bf v}+{\bf v_{\rm s}}\right)-\kappa \,{\bf n}\left({\bf n}\cdot\nabla E_c\right),
\label{equilibrium_flux}
\eeq
$E_c$ is the CR energy density, $p_c=E_c/3$, $\kappa$ is the diffusion coefficient, and ${\bf n}={\bf v_{\rm A}}/|\bf v_{\rm A}|$.   Equation \ref{cr_energy}, and indeed the scalar CR pressure in equation \ref{motion}, is formally valid only on scales larger than the mean-free-path of the $\sim$ GeV energy CRs that dominate the total energy of the CR population (e.g., \citealt{Skilling1971}).\footnote{Even given this restrictive assumption, the diffusion term in equation \ref{equilibrium_flux} in general depends on local plasma conditions.  It can also depend on the cosmic ray energy density itself, in which case the `diffusion' term in eq. \ref{equilibrium_flux} is not even truly diffusive (e.g., \citealt{Skilling1971,Wiener2013}).  We do not consider this complication in the present work.}

Throughout this paper we focus exclusively on the case of CR diffusion, which corresponds to ${\bf v_s}=0$ in equation \ref{equilibrium_flux}.   In a companion paper, we consider the case of CR streaming.  We further assume a hydrodynamic model which drops the field-aligned diffusive flux in equation \ref{equilibrium_flux} in favor of an effective isotropic diffusion equation with a spatially constant diffusion coefficient $\kappa$. We also consider a simple model for the gravity of a galaxy and its host dark matter potential: an isothermal sphere for which $\Phi = 2 V_g^2 \ln r$ where $\sqrt{2} V_g$ is the circular velocity of the potential.   Finally, we simplify the thermodynamics of the gas by using an isothermal equation of state with sound speed $c_i$, as would be appropriate for a warm gas in ionization equilibrium, or which might approximate the effects of  turbulence in the atmosphere of the host galaxy.  In addition, we consider a single phase flow and do not consider possible variation of CR transport with, e.g., the ionization state of the gas.

With these approximations, the steady state equations of motion are 
\be
\dot M_w = 4 \pi r^2 \rho v = {\rm const},
\label{eq:mdot}
\ee
\be
v \frac{dv}{dr} = -\frac{1}{\rho}\frac{dp}{dr} - \frac{1}{\rho}\frac{dp_c}{dr} - \frac{2 V_g^2}{r},
\label{eq:mom}
\ee
and 
\be
v\frac{dp_c}{dr} = -\frac{4 p_c}{3 r^2} \frac{d r^2 v}{dr} + \frac{1}{r^2}\frac{d}{dr}\left(r^2 \kappa \frac{d p_c}{dr}\right).
\label{eq:pc}
\ee

A key simplification can be made for the purposes of analytic estimates by noting that the order of magnitude of the diffusion term in equation \ref{eq:pc} relative to both of the other terms (which describe CR advection and adiabatic energy changes) is 
\be
\frac{\rm Diffusion}{\rm Advection} \sim \frac{\kappa}{r v} \sim  \left(\frac{\kappa}{10^{28} \, {\rm cm^2 \, s^{-1}}}\right) \left(\frac{30 \, {\rm km \, s^{-1} \, kpc}}{v \, r}\right),
\label{eq:dva}
\ee
which suggests that in the limit of rapid diffusion, and near the base of the outflow where the gas velocity is small, the advective and adiabatic terms in the CR energy equation can simply be neglected. For reference, the diffusion coefficient in the Milky Way is estimated to be $\sim 10^{29}$ cm$^2$ s$^{-1}$ (e.g., \citealt{Trotta2011}), although there is significant uncertainty in this estimate because of a degeneracy between the diffusion coefficient and the size of the CR propagation region (the `halo;' \citealt{Linden2010,Trotta2011}).   In \S \ref{sec:pion}, we return to constraints on the diffusion coefficient in galaxies using  observations of gamma-ray emission from pion decay. For now, we proceed under the assumption that for sufficiently large diffusion coefficients $\kappa$, one can neglect the advective and adiabatic terms in equation \ref{eq:pc} and focus solely on the diffusion term, for which the steady state solution is
\be
p_c = p_{c,0} + \frac{\dot E_c}{12 \pi \kappa} \left(\frac{1}{r} - \frac{1}{r_0}\right)
\label{eq:pcsol}
\ee
where $p_{c,0}$ is the base CR pressure at radius $r_0$ and 
\be 
\dot E_c = -12 \pi r^2 \kappa \frac{dp_c}{dr} = -12 \pi r_0^2 \kappa \frac{dp_c}{dr}\bigg|_{r_0} 
\label{eq:edotcdef}
\ee
is the energy per unit time supplied to the CRs by supernovae and other processes in the galaxy.  The assumption that diffusion is much faster than advection near the base of the wind implies that $\dot E_c$ is independent of radius.   In our numerical wind solutions in \S \ref{sec:numerics} we shall see that this assumption is valid at small radii near the sonic point (as assumed here to estimate $\dot M$), but that advection then takes over as the dominant energy transport mechanism at larger radii (see Fig. \ref{fig:Edot_diff} discussed below).

We will make frequent use of the CR scale height near the base of the wind:
\be
H_c = \left(\frac{- d \ln p_c}{d r}\bigg|_{r_0}\right)^{-1} \equiv r_0 \, h_c
\label{eq:Hc}
\ee
where we have defined $h_c$ as the dimensionless CR pressure scale-height at $r_0$.  With this definition, $\dot E_c = 12 \pi r_0 h_c^{-1} \kappa p_{c,0}$ and equation \ref{eq:pcsol} becomes $p_c = p_{c,0}[1 + h_c^{-1}(r_0/r-1)]$.   Note that solutions with $h_c < 1$ have $p_c \rightarrow 0$ at a finite radius $r=r_0/(1-h_c)$.  If CR diffusion were the only energy transport mechanism, $h_c \simeq 1$ would be the physical solution extending to large radii.  However, we will see below that $h_c \lesssim 1$ is typically the appropriate approximate solution at small radii (where our analysis here applies), because advection of CR energy eventually takes over from diffusion as the dominant transport mechanism.

Equations \ref{eq:mdot} and \ref{eq:mom} can be combined to yield a  wind equation
\be
\frac{1}{v}\frac{dv}{dr}\left(v^2 - c_i^2 \right) = \frac{2 c_i^2}{r} -\frac{1}{\rho}\frac{dp_c}{dr} - \frac{2 V_g^2}{r}.
\label{eq:wind1}
\ee
With equation \ref{eq:pcsol}, i.e., assuming $\dot E_c$ is a constant because diffusion is rapid, the wind equation becomes
\be
\frac{1}{v}\frac{dv}{dr}\left(v^2 - c_i^2 \right) = \frac{2 c_i^2}{r} + \frac{\dot E_c}{12 \pi r^2 \rho \kappa} - \frac{2 V_g^2}{r}.
\label{eq:wind}
\ee

In the portion of the flow where $v < c_i$, an approximation to the density profile can be derived by assuming hydrostatic equilibrium so that
\be
\frac{c_i^2}{\rho}\frac{d\rho}{dr} - \frac{\dot E_c}{12 \pi r^2 \rho \kappa} = - \frac{2 V_g^2}{r}.
\label{eq:HE}
\ee
Equation \ref{eq:HE} has the solution
\be
\rho(r) = \rho_0 \left[x^{-\xi} + A(x^{-1} - x^{-\xi})\right]
\label{eq:rho}
\ee
where $x = r/r_0$ and $\xi$ and $A$ are the two key dimensionless parameters in this problem, which measure the strength of gravity relative to the gas and CR pressures at the outflow's base: 
\be
\xi \equiv \frac{2 V_g^2}{c_i^2}
\label{eq:xi}
\ee
and
\be \begin{split}
A & \equiv  \frac{\dot E_c}{12 \pi \kappa  c_i^2 \rho_0 r_0 (\xi - 1)} \simeq \frac{\dot E_c}{24 \pi \kappa V_g^2 \rho_0 r_0} \\ 
& \simeq \frac{p_{c,0}}{2 h_c \rho_0 V_g^2} = \frac{1}{2 h_c}\frac{c_{c,0}^2}{V_g^2}\lesssim 1,
\label{eq:A}
\end{split}
\ee
where we assume $\xi \gg 1$ in the approximations after the first equality, and where the last equality defines the base CR sound speed $c_{c,0}=(p_{c,0}/\rho_0)^{1/2}$. The conclusion that $A \lesssim 1$ follows from the fact that $A$ is roughly the ratio of the CR pressure force to the gravitational force at the base of the outflow. If that ratio is $\gtrsim 1$, then the initial assumption of HE is invalid and the gas distribution would expand out until $A \lesssim 1$.   

The two power-laws in equation \ref{eq:rho}, namely $\rho \sim r^{-\xi}$ and $\rho \sim r^{-1}$, correspond to the gas pressure and CR pressure dominated phases of the solution, respectively.   The transition between the two happens at a radius $r_{tr} \simeq r_0 \, A^{-1/(\xi-1)}$.  The CR-dominated hydrostatic $\rho \propto p_{c,0}/r$ solution in eq. \ref{eq:rho} was derived by \citealt{Ji2020} for the CGM in their cosmological zoom-in simulations with CRs.   \citet{Hopkins2021} further used this solution to estimate the properties outflows on CGM scales driven by CRs.  The key difference between the solutions here and their estimates is that we self-consistently match our CR-dominated solution onto the gas-pressure dominated solution near the galaxy in order to calculate the properties of galaxy-scale winds.   We do not include a CGM at larger radii in our calculations.  

The mass-loss rate in the wind is set by the conditions at the critical (sonic) point of equation \ref{eq:wind}, which we denote as $r_s$. Setting the numerator and denominator of the wind equation to zero simultaneously yields two conditions: 
\be
v(r_s) = c_i \ \ \ {\rm and} \ \ \ \frac{r_s \, \rho(r_s)}{r_0 \rho_0} = \frac{c_{c,0}^2}{2 h_c \Veffsq}.
\label{eq:critical}
\ee
where
\be
\Veffsq=V_g^2-c_i^2,
\ee
These conditions allow for a compact expression for the wind mass-loss rate:
\be
\dot M_w = 4 \pi r_s^2 \rho(r_s) c_i = \frac{2 \pi r_s r_0 \rho_0 c_i}{h_c} \frac{c_{c,0}^2}{\Veffsq}.
\label{eq:firstmdot}
\ee
To estimate the radius of the sonic point $r_s$, one can numerically solving the 2nd critical point equation in eq. \ref{eq:critical} using the analytic density profile in equation \ref{eq:HE}, i.e.,
\be
\frac{c_{c,0}^2}{2 h_c (V_g^2-c_i^2)} \simeq x_s^{-\xi+1} + \frac{c_{c,0}^2}{h_c (2V_g^2-c_i^2)}\left(1 - x_s^{-\xi+1}\right)
\label{eq:rcnum}
\ee
A good approximate solution to equation \ref{eq:rcnum} can be found by expanding for $c_i^2 \ll 2 V_g^2$ which yields
\be
r_{s} \simeq r_0 \, \left(\frac{4 h_c V_g^4}{c_{c,0}^2 c_i^2}\right)^{c_i^2/(2V_g^2)}
\label{eq:rc}
\ee
As an example, if $c_i \simeq c_{c,0} \simeq 10 \kms$ (as in the Milky Way), equation \ref{eq:rc} predicts $r_s/r_0 \simeq 1.38,$ 1.13, and 1.05 for $h_c=1$ and $V_g = 30$, 60, and 100\,$\kms$, while numerical solution of equation \ref{eq:rcnum} yields nearly identical results of $r_s/r_0 \simeq 1.39$, 1.13, and 1.05, respectively.    The sonic point is quite close to the base of the wind, i.e., $r_s \simeq r_0$, unless $c_i \sim V_g$.  

Using equation (\ref{eq:rc}) for the sonic point in equation (\ref{eq:firstmdot}), we obtain an expression for the mass-loss rate:  
\be
\dot{M_w}\simeq \frac{2 \pi r_0^2\rho_0 c_i}{h_c} \, \frac{c_{c,0}^2}{\Veffsq} \left(\frac{4 h_c V_g^4}{c_{c,0}^2 c_i^2}\right)^{c_i^2/(2V_g^2)}.
\label{eq:mdotd}
\ee
In the limit that $2V_g\gg c_i$,
\be
\dot M_w \sim \frac{2 \pi r_0^2 c_i p_{c,0}}{h_c V_g^2}=\frac{2 \pi r_0^2 \rho_0 c_i}{h_c} \left(\frac{c_{c,0}}{V_g}\right)^2.
\label{eq:mdot2}
\ee
We stress that for equation \ref{eq:mdot2} to be applicable, CR diffusion must  dominate over advection out to at least the sonic point since the mass-loss rate is set by the flow properties at and interior to the sonic point.  This, together with equation \ref{eq:dva}, implies that the relevant criterion for the validity of our analytics is roughly $\kappa \gtrsim r_0 c_i$.   We shall see that this is borne out by the numerical simulations in \S \ref{sec:numerics}.  Diffusion coefficients satisfying this constraint are also strongly suggested by gamma-ray data on pion decay in nearby star-forming galaxies, as we show in \S \ref{sec:pion}.

Note that equations \ref{eq:mdotd} and \ref{eq:mdot2} imply that the mass-loss rate does not explicitly depend on the diffusion coefficient $\kappa$ at fixed base CR pressure $p_{c,0}$ (though there is an implicit dependence via $h_c$ as we will see below).  However, for a fixed CR injection power $\dot E_c$, equations \ref{eq:mdotd} and  \ref{eq:mdot2} imply $\dot M_w \propto 1/\kappa$ because the base CR pressure itself scales as $p_{c,0}=\rho_0 c_{c,\,0}^2 \propto \dot E_c/\kappa$ (see eq. \ref{eq:pcsol}). Substituting into equation (\ref{eq:mdotd}), we find that 
\be
\dot{M}_w\simeq\,\left(\frac{r_0\,c_i}{6\kappa}\right)\left(\frac{\dot{E}_c}{\Veffsq}\right)\left(\frac{4 h_c V_g^4}{c_{c,0}^2 c_i^2}\right)^{c_i^2/(2V_g^2)},
\label{eq:mdotedot}
\ee
which makes the $\kappa$ dependence, and the competition between diffusion and advection at the base of the outflow, explicit.

We now present an order of magnitude estimate of the scale-height $h_c$ by determining the radius $r_{\rm adv} \simeq r_0(1 + h_c)$ at which the advective flux $F_{\rm adv} = 4 p_c v = \dot M_w p_c/(\pi r^2 \rho)$ is comparable to the diffusive flux $F_{\rm diff} = - 3 \kappa d p_c/dr$. This occurs when the velocity of the outflow is given by
\be
v(r_{\rm adv}) \approx - \kappa \frac{d \ln p_c}{dr} \approx \frac{c_i}{h_c} \frac{\kappa}{c_i r_0} 
\label{eq:radv}
\ee
where we have dropped a factor of $3/4$ consistent with the rough nature of the estimates that follow.    Since $h_c \lesssim 1$ and our analytics assumes $\kappa \gtrsim c_i r_0$, equation \ref{eq:radv} implies that the transition from diffusion to advection happens exterior to the sonic point.  Gas pressure is then negligible and the momentum equation becomes $\rho v dv/dr \simeq -dp_c/dr - 2\rho V_g^2/r$.   At the sonic point, the two terms on the right-hand-side of $dv/dr$ are comparable (see eq. \ref{eq:critical}).  Since the gas density scale-height is smaller than the CR scale-height, somewhat exterior to the sonic point $\rho v dv/dr \simeq -dp_c/dr$. Multiplying by $4 \pi r^2$ gives 
\be
\dot M \frac{dv}{dr}\bigg|_{\sim r_{\rm adv}} \sim 4 \pi r_{\rm adv}^2 \frac{p_{c,0}}{h_c} \rightarrow v(r_{\rm adv}) \sim \frac{h_c V_g^2}{c_i}
\label{eq:radv2}
\ee
where in the second expression we have used the approximate version of $\dot M$ from equation \ref{eq:mdot2} and have taken $r_{\rm adv} \approx r_0$, consistent with $h_c \lesssim 1$.   Equating equations \ref{eq:radv} and \ref{eq:radv2} then yields $h_c \sim (c_i/V_g)(\kappa/r_0 c_i)^{1/2}$.   For $\kappa \rightarrow \infty$ this gives $h_c >> 1$, inconsistent with the local approximation used here;  the solution should be $p_c \propto 1/r$, i.e., $h_c \simeq 1$, so that
\be
h_c \sim \min\bigg(1,\frac{c_i}{V_g}\sqrt{\frac{\kappa}{r_0 c_i}}\bigg).
\label{eq:hc}
\ee
As an example, if $\kappa \sim 10^{29}$ cm$^2$ s$^{-1} \simeq 10 \, (r_0 c_i/30 \kms \, {\rm kpc})$, $c_i \sim 10 \, \kms$, and $V_g \sim 100 \, \kms$, $h_c \sim 1/3$.    In our analytic scalings that follow in this section we primarily normalize $h_c$ to a value of $1/4$ motivated by these fiducial parameters, but in several plots we will use equation \ref{eq:hc}.  In \S \ref{sec:numerics}, we also compare equation \ref{eq:hc} to our numerical solutions and find good agreement. 

Our final expression for the mass-loss rate in equation \ref{eq:mdotd} can be written as
\be
\begin{split}
\dot M_w \simeq & \frac{4 \pi r_0^2 \,\rho_0\, c_i}{h_c}\,\left(\frac{ c_{c,0}^2}{2\Veffsq}\right)\left(\frac{4 h_c V_g^4}{c_{c,0}^2 c_i^2}\right)^{c_i^2/(2V_g^2)}\\
 \simeq & \,\, 0.06 \, \mspy \left(\frac{r_0}{1 \, \kpc}\right)^2 \left(\frac{1/4}{h_c}\right) \left(\frac{4 h_c V_g^4}{c_{c,0}^2 c_i^2}\right)^{c_i^2/(2V_g^2)}\\ &
\times \, 
\left(\frac{n_0}{1 \, {\rm cm^{-3}}}\right)   \left(\frac{c_i \ c^2_{c,0}}{[10 \, \kms]^3}\right) \left(\frac{\Veff}{100 \, \kms}\right)^{-2} 
\label{eq:mdot3}
\end{split}
\ee
where $n_0=\rho_0/m_p$.   For reference, if we scale for parameters appropriate to a galaxy like the Milky Way (with $V_g\simeq150$\,km s$^{-1}$, $c_i = c_{c,0}\simeq10$\,km s$^{-1}$, $n_0\simeq 1$\, cm$^{-3}$, and $r_0\sim5$\,kpc), equation \ref{eq:mdot3} yields $\dot M_w \simeq 1 \mspy$, comparable to the star formation rate.  Equation \ref{eq:mdot3} also predicts $\dot M_w \propto p_{c,0} c_i/h_c$, which scales $\propto p_{c,0} c_i$ or $\propto p_{c,0} c_i^{1/2}$, depending on which regime of equation \ref{eq:hc} is appropriate (these scalings assume $r_s \simeq r_0$ for simplicity). To the (uncertain) extent that the CR pressure is comparable in different phases of the ISM, the outflow is thus likely to be dominated by the warmer ISM phases, though only by a factor of a few.

\begin{figure}
\centering
\includegraphics[width=84mm]{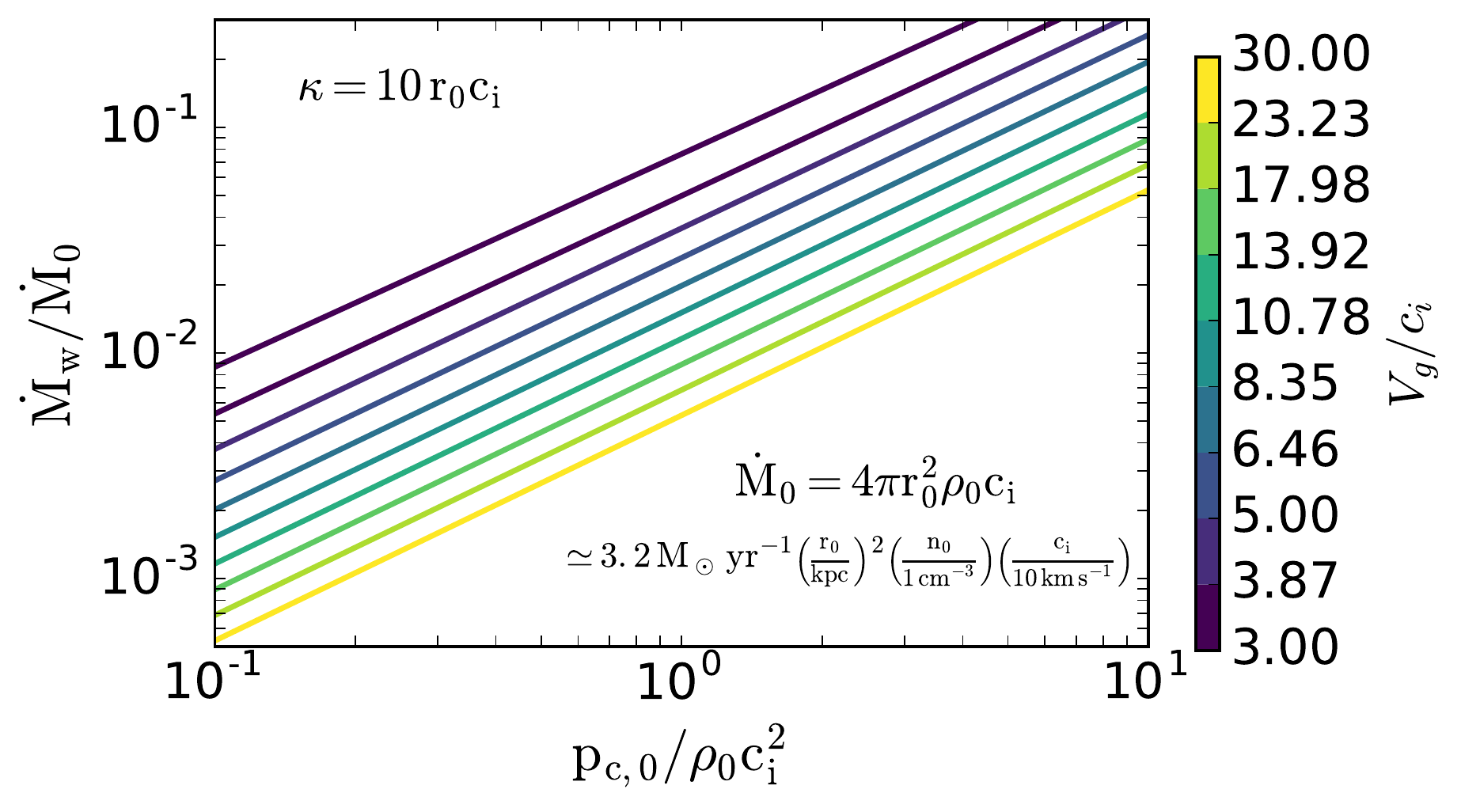}
\vspace{-0.2cm}
\caption{Analytic mass-loss rates for CR driven galactic winds in the limit of rapid diffusion (eq. \ref{eq:mdot3} with $h_c$ from eq. \ref{eq:hc}), as a function of the strength of gravity relative to the gas sound speed in the disc ($V_g/c_i$) and the base CR pressure (${\rm p_{c,0}/\rho_0 c_i^2}$). The mass-loss rates are normalized by equation (\ref{eq:mdot0}). Labeled values of $V_g/c_i$ on the color bar are logarithmically distributed and correspond to the curves on the plot.} 
\label{fig:mdotdiff}
\end{figure}

Figure \ref{fig:mdotdiff}  shows the mass-loss rate (eq. \ref{eq:mdot3} with $h_c$ from eq. \ref{eq:hc} taking $\kappa = 10 r_0 c_i$) as a function of the two key dimensionless parameters in the problem, namely the strength of gravity ($V_g/c_i=(\xi/2)^{1/2}$; see eq.~\ref{eq:xi}) and the base CR sound speed  ($p_{c,0}/\rho_0 c_i^2=c_{c,0}^2/c_i^2\propto A/\xi$); see eq.~\ref{eq:A}).   The mass-loss rate is given in units of
\be
\dot M_0 = 4 \pi r_0^2 \rho_0 c_i \simeq 3.2 \, \mspy \,  \left(\frac{r_0}{\kpc}\right)^2 \left(\frac{n_0}{1 \, {\rm cm^{-3}}}\right) \left(\frac{c_i}{10 \, \kms}\right).
\label{eq:mdot0}
\ee

To express the mass-loss rate in CR driven winds in a more intuitive form we take $V_g \gg c_i$ and use the simplified expression for the mass-loss rate in equation \ref{eq:mdot2}. We then write the CR energy injection rate at the base of the outflow as 
\be
\dot E_c = \epsilon_c \dot M_* c^2,
\label{eq:edotc}
\ee
where $\dot{M}_*$ is the star formation rate and $\epsilon_c \equiv 10^{-6.3} \epsilon_{c,-6.3}$ is related to the fraction of SNe energy that goes into CRs:  for $10^{51}$\,ergs per SNe and 1 SNe per 100 $M_\odot$ of stars formed, $\epsilon_c = 10^{-6.3}$ if $10 \%$ of the SNe energy goes into primary CRs.  Equation \ref{eq:mdot2} can then be approximately rewritten as (taking $r_s \simeq r_0$ to simplify eq. \ref{eq:rc})
\begin{eqnarray} 
\frac{\dot M_w}{\dot M_*} &\simeq& 0.8 \ \epsilon_{c,-6.3}  \left(\frac{c_i r_0}{\kappa}\right) \left(\frac{100 \, \kms}{V_{\rm g,\,eff}}\right)^{2}\nonumber \\
&\simeq& 0.08 \ \epsilon_{c,-6.3} \, \left(\frac{c_i r_0}{{30\,\rm km\,\,s^{-1} \, kpc}}\right) \nonumber \\ & &\times \left(\frac{10^{29}\,{\rm cm^2\,s^{-1}}}{\kappa}\right) \left(\frac{100 \, \kms}{V_{\rm g,\, eff}}\right)^{2}.
\label{eq:mdotf}
\end{eqnarray}
Per equation \ref{eq:dva} and the associated discussion, $\kappa \sim \, 10 c_i r_0$ is plausible; we have scaled to representative values in the second line. This expression shows that significant wind mass loading relative to the global star formation rate is in principle possible, and that it should grow strongly with decreasing $V_{\rm g,\,eff}$.  However, if $\kappa \sim 10^{29}$ cm$^2$ s$^{-1}$ is appropriate, the mass-loading factors in dwarf galaxies are relatively modest compared to the values $\dot M_w \gg \dot M_*$ needed to reconcile the stellar and dark matter mass functions.   We return in Section \ref{sec:pion} to an observational calibration of the diffusion coefficients in other star-forming galaxies.

A second instructive expression for the wind mass-loss rate  can be obtained by comparing the mass-loss rate estimated here to the star formation rate predicted by feedback-regulated models of star formation in galaxies, namely $\dot M_* \approx \pi r_0^2 \dot \Sigma_*$ where the surface density of star formation is (e.g., \citealt{Thompson2005})
\be
\dot \Sigma_* \simeq \frac{\sqrt{8} \pi G \Sigma_g^2 \phi}{v_*} 
\label{eq:SFR}
\ee
where $\phi = 1 + \Sigma_{*}/\Sigma_g$ describes the contribution of the stellar disc with surface density $\Sigma_*$ to the local gravitational potential and $v_* = p_*/m_* \approx 3000 \kms$ is the momentum per unit mass of star formed associated with stellar feedback, which supports the disc against its own self-gravity \citep{Ostriker2011}. Equation \ref{eq:mdot2} for the mass-loss rate can then be recast as
\be
\frac{\dot M_{\rm w}}{\dot M_*} \simeq 8\left(\frac{1/4}{h_c}\right) \left(\frac{c_i v_*}{3 \times 10^4 \, {\rm km^2 \, s^{-2}}}\right) \left(\frac{100 \, \kms}{V_g}\right)^{2} \frac{p_{c,0}}{\pi G \Sigma_g^2 \phi}
\label{eq:mdotdf2}
\ee
Equation \ref{eq:mdotdf2} again shows that for Milky-way like conditions in which the CR pressure is comparable to that needed for hydrostatic equilibrium in the galactic disc ($p_{c,0}\simeq 1/3 \, \pi G\Sigma_g^2 \phi$; \citealt{Boulares1990}), the wind mass-loss rate driven by diffusing CRs can be of order or larger than the star formation rate. As we discuss in Section \ref{sec:pion}, equation (\ref{eq:mdotdf2}) also shows that for dense starburst galaxies, in which $p_{c,0}\ll \pi G\Sigma_g^2 \phi$ \citep{Lacki2010,Lacki2011}, the mass-loss rate in CR driven winds is significantly reduced.

\subsection{Wind Energetics, Momentum Flux, and Velocity}
 
For a steady state solution, the momentum equation (eq. \ref{eq:mom}) and CR energy equation with diffusion (eq. \ref{eq:pc}) can  be combined to yield a total conserved energy outflow rate, namely
\be
\dot E_{w} = \dot M_w \left(\frac{1}{2} v^2 + c_i^2 \ln \rho + 4 c_c^2 + \Phi \right) + \dot E_{\rm c} = {\rm constant}
\label{eq:Edotdiff}
\ee
where $\dot E_{\rm c} = - 12 \pi r^2 \kappa \,dp_c/dr$ (as before) and where the four terms in parentheses in equation \ref{eq:Edotdiff} correspond to the gas kinetic energy flux ($\equiv \dot E_k$), the gas enthalpy/advective flux (assuming our isothermal equation of state for the gas), the CR enthalpy/advective flux ($\equiv \dot E_{c,\,\rm adv}$), and the gravitational energy flux ($\equiv \dot E_{\rm grav}$), respectively.   Note that equation \ref{eq:Edotdiff} no longer assumes that diffusion dominates over advection at all radii.

\begin{table*}
\caption{Parameters and properties of our numerical simulations.   Numerical resolution is $\delta r/r = 5.25 \times 10^{-4}$ unless otherwise noted.  Columns are diffusion coefficient $\kappa$, velocity $V_g$ of the isothermal gravitational potential, base CR pressure $p_{c,0}$, outer radius of domain $r_{out}$, reduced speed of light $V_m$, simulation mass0loss rate $\dot M_{sim}$ in units of $\dot M_0$ (eq. \ref{eq:mdot0}), simulation mass-loss rate in units of $\dot M_{max}$ (the maximum mass-loss rate allowed by energy conservation; eq. \ref{eq:mdotmaxsim}), the dimensionless base CR scale-height $h_c$ (eq. \ref{eq:Hc}), and the kinetic energy and CR enthalpy fluxes at the top of the domain relative to the diffusive CR flux at the base $(\dot E_k(r_{out}) + \dot E_{c,adv}(r_{out}))/\dot E_c(r_0)$.   When the latter is $\simeq 1$, most of the CR energy flux at the base of the wind goes into kinetic energy at large radii, while when it is $\lesssim 1$, most of the CR energy goes into work against gravity, and $\dot M_w \simeq \dot M_{max}$ (eq. \ref{eq:mdotmax} \& \ref{eq:mdotmaxsim}).}
\begin{tabular}{cc|ccccccccccc}
  $\kappa$ & $V_g$ & $p_{c,0}$ & $r_{out}$ & $V_m$ & $\dot M_{sim}$ &  $\dot M_{sim}$ & $v(r_{out})$ & $h_c$ & $\dot E_k(r_{out}) + \dot E_{c,adv}(r_{out})$  \\ [1pt]
  ($r_0 c_i$) & ($c_i$) & ($\rho_0 c_i^2$) & ($r_0$) & ($c_i$) & ($\dot M_0$) & ($\dot M_{max}$) & ($V_g$) & -- & $ (\dot E_c[r_0])$ \\ [3pt] \hline
    $ 20^{a}$ &  200 & 400 & 7.6  & 60000 & 0.21  & 0.032 & 13.8 & 0.023 & 0.95  \\
    $ 10^{b} $ &  100 & 100 & 7.6  & 30000 & 0.16 & 0.067 & 9.5 & 0.03 & 0.94    \\
    $ 3.3^{c} $ &  33 & 11 & 7.6  & 10000 & 0.09 & 0.19 & 5 & 0.053 & 0.8  \\
    $ 30 $ &  10 & 1 & 5  & 3000 & 0.016 & 0.018 & 13.6 & 0.32 & 0.93 \\
   $ 10 $ &  10 & 1 & 5  & 3000 & 0.022  & 0.054  & 8.2 & 0.23 & 0.93 \\
    $ 3.3$ &  10 & 1 & 5  & 3000  & 0.03 & 0.15 & 4.6  & 0.16 & 0.83 \\
    $ 2$ &  10 & 1 & 5  & 3000  & 0.03 & 0.42 & 3.3  & 0.13 & 0.57 \\
    $ 1 $ &  10 & 1 & 15  & 3000  & 0.03 & 0.75 & 0.5 & 0.14 & 0.23 \\
     $ 0.33 $ &  10 & 1 & 15 & 3000  & 0.007 & 0.78 & 0.4 & 0.21 & 0.16  \\
     $ 0.11 $ & 10 & 1 & 25 & 3000  & 0.0019 & 0.85 & 0.43 & 0.23 & 0.08 \\
   $ 10 $ & 10 & 0.3 & 5  & 3000  & 0.007 & 0.058 & 8.2 & 0.23 & 0.99 \\
     $ 10 $ &  10 & 0.1 & 5 & 3000 & 0.0027 & 0.056 & 8.1 & 0.24 & 0.95 \\
   $ 10 $ &  10 & 3 & 15  & 3000 & 0.065 & 0.089 & 9.4 & 0.23 & 0.9  \\
     $ 10 $ &  6 & 1 &  5  & 3000 & 0.043 & 0.055 & 7.5 & 0.33 & 0.89 \\
    $ 3.3 $ &  6 & 1 &  15  & 3000 & 0.055 & 0.26 & 4.9 & 0.24 & 0.67 \\
   $ 10 $ &  3 & 1 & 15  & 3000 & 0.136 & 0.11 & 7.6 & 0.5  & 0.9 \\
   $ 3.3 $ &  3 & 1 & 15  & 3000 & 0.156 & 0.3 & 4.2 & 0.4 & 0.74 \\    [3pt] \hline
\label{tab:compare}
\end{tabular}
\\ 
$^{a}$ $\delta r/r=8.25\times 10^{-6}$ for $1\leq r \leq 1.14$; $c_i$ is 1/20 of $\kappa = 1$, $V_g=10$ sim \\
$^{b}$ $\delta r/r=2.5\times 10^{-5}$ for $1\leq r \leq 1.11$; $c_i$ is 1/10 of $\kappa = 1$, $V_g=10$ sim \\
$^{c}$ $\delta r/r=6.6\times 10^{-5}$ for $1\leq r \leq 1.11$; $c_i$ is 1/3 of $\kappa = 1$, $V_g=10$ sim \\
\end{table*}

The total energy flux and terminal velocity of the wind can be estimated as follows.  Under our assumption of $\kappa \gtrsim r_0 c_i$, at small radii near the base of the wind, the energy flux at small radii is almost entirely due to CR diffusion, so that $\dot E_{w} \simeq \dot E_{\rm c}$.  This neglects gravity near the base of the wind, which we return to below.   The CR diffusive flux near the base can be related to the mass-loss flux using equation \ref{eq:mdotedot} where in this expression we now identify $\dot E_c$ near the base as the total wind luminosity $\dot E_w$ in equation \ref{eq:Edotdiff}.   At large radii, the energy flux will be dominated by the gas with $\dot E_w \simeq 0.5 \dot M_w v_\infty^2$.   Equating our expressions for the total wind luminosity at small and large radii yields the terminal velocity of the wind
\be
\begin{split}
& v_\infty \simeq  2 \Veff \left(\frac{3 \kappa}{r_0 c_i}\right)^{1/2} \left(\frac{4 h_c V_g^4}{c_{c,0}^2 c_i^2}\right)^{-c_i^2/(4V_g^2)}
\label{eq:vinfdiff}
 \\ & \simeq 10^3 \, \kms \left(\frac{V_g}{100 \, {\rm km \,s^{-1}}}\right) \left(\frac{\kappa}{10^{29} {\rm cm^2 \,s^{-1}}}\frac{\rm 30  \, kpc \, {\rm km \,s^{-1}}}{r_0 c_i}\right)^{1/2}
\end{split}
\ee
where the second expression assumes $V_g \gg c_i$.  Recall that we require $\kappa \gtrsim r_0 c_i$ given our assumptions used in deriving the approximate wind solutions here.  In this case, equation \ref{eq:vinfdiff} shows that $v_\infty \gtrsim \Veff$, with $v_\infty \sim 1-10 V_g$ plausible.   We show below (Fig. \ref{fig:mdotfin}) that equation \ref{eq:vinfdiff} agrees very well with our numerical simulations.

Using equations (\ref{eq:mdot3}) and (\ref{eq:vinfdiff}) we can write down an expression for the asymptotic total momentum loss rate carried by the wind, 
\be
\dot{p}_w=\dot{M}_w v_\infty
\simeq\frac{4\pi r_0^2 c_i p_{c,0}}{h_c V_{\rm g}}\left(\frac{3\kappa}{r_0c_i}\right)^{1/2}.
\ee
where we assume $V_g \gg c_i$ to simplify the expressions and highlight the key scalings.  Assuming no pionic losses in the host galaxy, we can write this expression in terms of the star formation rate using equations (\ref{eq:mdotedot}) \& (\ref{eq:edotc}) as
\be
\dot{p}_w=\dot{M}_w v_\infty
\simeq\frac{\epsilon_c\dot{M}_* c^2}{V_{\rm g}}\left(\frac{r_0 c_i}{3\kappa}\right)^{1/2}.
\ee
This quantity can be compared with the total momentum rate carried by photons from star formation $\dot{p}_{*}=L_{*}/c=\epsilon_{*}\dot{M}_* c$, where $\epsilon_{*,\,-3.3}=\epsilon_{*}/5\times10^{-4}$ for steady-state star formation and a standard IMF. We then have that 
\begin{eqnarray}
\frac{\dot{p}_w}{\dot{p}_{*}}&\simeq&\frac{\epsilon_c}{\epsilon_{*}}\frac{c}{V_{\rm g,\,eff}}\left(\frac{r_0 c_i}{3\kappa}\right)^{1/2} \nonumber \\
&\simeq&0.5\,\frac{\epsilon_{c,-6.3}}{\epsilon_{*,\,-3.3}}\left(\frac{100\,\rm km \,\,s^{-1}}{V_{\rm g}}\right) \nonumber \\
&\times&\left(\frac{r_0\,c_i}{30 \,\rm km \,\,s^{-1}\,\, kpc}\frac{\rm 10^{29}\,cm^2\,\,s^{-1}}{\kappa}\right)^{1/2}.
\end{eqnarray}
Note that the asymptotic kinetic energy loss rate $\dot E_w = 0.5\dot{M}_w v_\infty^2 = \dot E_c = \epsilon_c\dot{M}_* c^2$.   CR-driven winds in the rapid diffusion approximation are thus energy-conserving in the sense that the wind kinetic energy power at large radii is of order that supplied to the CRs at small radii.  As in the discussions following equations (\ref{eq:mdotf}) and (\ref{eq:mdotdf2}), we reiterate that this conclusion assumes that there are no pionic losses in the host galaxy (see \S\ref{sec:pion}). 

\begin{figure*}
\centering
\includegraphics[width=174mm]{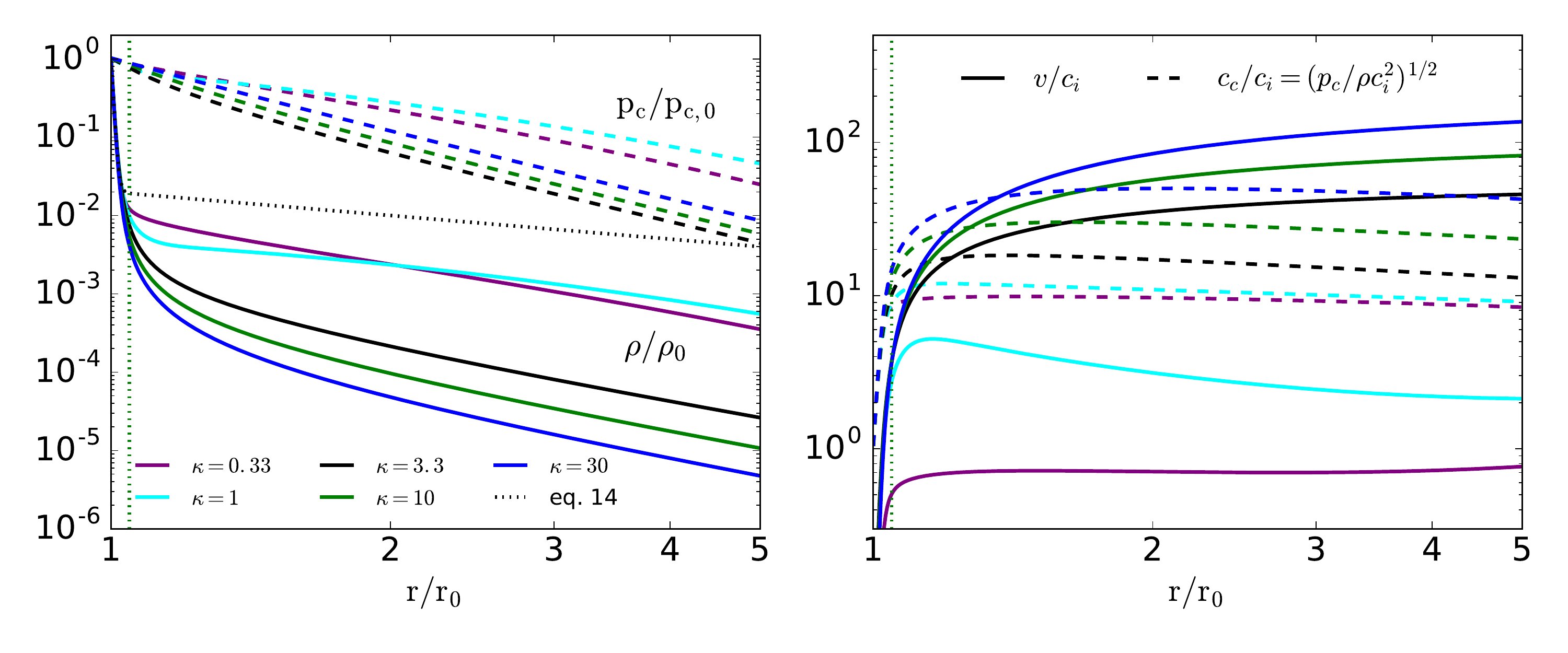}
\vspace{-0.2cm}
\caption{{\em Left:} CR pressure and gas density as a function of radius for $V_g=10$ and several  diffusion coefficients (in units of $r_0 c_i$; see Table \ref{tab:units}).  The dotted profile is the analytic hydrostatic solution in equation \ref{eq:rho} for $h_c=1/4$. The analytic density profile is reasonable interior to the sonic point (and thus for estimating $\dot M$) but not at larger radii for the higher $\kappa$ solutions.   {\em Right:}  Flow velocity and CR sound speed for the same solutions, in units of gas sound speed $c_i$.   The vertical green dotted line in both panels is the analytic estimate of the location of the critical point (eq. \ref{eq:rc}).  For the $\kappa=3-30$ solutions, the flow accelerates very rapidly and the velocity exceeds the escape velocity just exterior to the base of the wind.  For $\kappa = 1$ the flow is significantly slower, and the density profile is closer to hydrostatic.  For $\kappa=0.33$ the velocities are subsonic and well below the escape speed at all radii shown here.  We return to the $\kappa = 0.33$ solution in Figure \ref{fig:Edot_diff_lowkap} and show that it does drive a wind but one whose properties are very different from the high $\kappa$ simulations.}
\label{fig:flow_diff}
\end{figure*}

 \subsection{Maximum Mass-Loss Rate}

There is a strict upper limit to the mass-loss rate associated with CR-driven winds that is set by energy conservation.  This is the regime in which gravity significantly modifies the energetics of the outflow and cannot be neglected as we did in deriving equation \ref{eq:vinfdiff}.  If the asymptotic speed of the wind vanishes then $\dot E_{c}(r_0) + \dot M_{\rm max} \Phi(r_0) \simeq 0$, i.e., $\dot M_{\rm max} \simeq 2 \dot E_{c}(r_0)/v^2_{\rm esc}(r_0)$ where $v_{\rm esc}(r_0)$ is the escape speed from the base of the wind.\footnote{$v_{\rm esc}(r_0)$ is formally not defined for the $\ln(r)$ potential focused on here, but it is $\sim 2 V_g$ for a more realistic potential which deviates from isothermal at larger radii (and/or for a computational domain of reasonable size even with a $\ln(r)$ potential).}  This regime is analogous to photon-tired winds in stellar wind theory \citep{Owocki1997}.  Using $\dot E_c = \epsilon_c \dot M_* c^2$ (eq. \ref{eq:edotc}), we can write
\be
\frac{\dot M_{\rm max}}{\dot M_*} \simeq \, 9 \, \epsilon_{c,-6.3} \, \left(\frac{100 \, \kms}{v_{\rm esc}(r_0)}\right)^2,
\label{eq:mdotmaxvsmdotstar}
\ee
which has an ``energy-driven" velocity scaling \citep{Murray2005}.   Alternatively, using $\dot E_{c}(r_0) \simeq 12 \pi r_0 \kappa p_{c,0}/h_c$ we find
\begin{eqnarray}
\dot{M}_{\rm max}&\simeq&\frac{24\pi r_0\kappa p_{c,\,0}}{v_{\rm esc}^2(r_0)h_c}\simeq\frac{24\pi r_0\kappa\rho_0}{h_c}\left(\frac{c_{c,\,0}}{v_{\rm esc}(r_0)}\right)^2 \nonumber \\
&\simeq&23\,{\rm M_\odot\,\,yr^{-1}}\left(\frac{r_0}{\rm kpc}\right)\left(\frac{\kappa}{10^{29}\rm cm^2\,\,s^{-1}}\right)\left(\frac{n_0}{\rm cm^{-3}}\right) \nonumber \\
&\times&\left(\frac{1/4}{h_c}\right)
\left(\frac{c_{c,\,0}}{10\,{\rm km\,\,s^{-1}}}\right)^2\left(\frac{100\,{\rm km\,\,s^{-1}}}{v_{\rm esc}(r_0)}\right)^2
\label{eq:mdotmax}
\end{eqnarray}
The maximum cosmic ray-driven wind momentum will be realized for mass-loss rates a bit below $\dot M_{\rm max}$ when $v_\infty \sim v_{\rm esc}(r_0)$ (at $\dot M_{\rm max}$, $v_\infty \equiv 0$ and so $\dot p_w = 0$).   A rough estimate is 
\begin{equation}
    \frac{\dot{p}_{w,\,{\rm max}}}{\dot{p}_*}\sim 5 \frac{\epsilon_{c,-6.3}}{\epsilon_{*,\,-3.3}} \left(\frac{100\,\rm km \,\,s^{-1}}{v_{\rm esc}(r_0)}\right).
\end{equation}
but more detailed calculations are required to precisely determine the numerical pre-factor in this expression.

It is straightforward to show that the mass-loss rate estimated in equation \ref{eq:mdot3} is 
\be
\dot M_w \simeq \dot M_{\rm max}\,\left(\frac{r_0 c_i}{3 \kappa}\right),  \label{eq:mdotwvsmdotmax}
\ee
where we have assumed $v^2_{\rm esc}(r_0) \simeq 4 V_g^2$.   Equation \ref{eq:mdotwvsmdotmax} shows that the wind mass-loss rate is in general $< \dot M_{\rm max}$ given our restriction to $\kappa \gtrsim r_0 c_i$.   Our analysis so far does not preclude that winds exist for $\kappa$ smaller than $r_0 c_i$, but our analytic approximations break down in this regime. Equation \ref{eq:mdotwvsmdotmax} shows, however, that if one extrapolates our estimate of $\dot M_w$ to $\kappa < r_0 c_i$, the solutions reach the maximum mass-loss rate allowed by energy conservation for $\kappa \sim 1/3 r_0 c_i$.   Thus a plausible conjecture is that any wind with $\kappa \lesssim r_0 c_i$ will be a CR analogue of photon-tired winds \citep{Owocki1997}.    In the next section we show that this conjecture is correct:   numerical simulations with $\kappa \lesssim r_0 c_i$ do produce a wind, but one whose character is very different from the high $\kappa$ simulations. See Section \ref{sec:lowkapnum}.  In particular, winds with $\kappa \lesssim r_0 c_i$ have $\dot M_w \simeq \dot M_{\rm max}$ (and thus lose most of their energy to work against gravity) and accelerate much more slowly, with the sonic point at radii many times larger than $r_0$.  Finally, we note that for $\kappa \equiv 0$, there cannot be a wind so long as the base CR and gas sound speeds are below the escape speed (as assumed here) since both fluids are then adiabatic.

Equations \ref{eq:mdotf}, \ref{eq:mdotmaxvsmdotstar}, \& \ref{eq:mdotwvsmdotmax} highlight the importance of the value of the diffusion coefficient for assessing the implications of our results for CR-driven galactic winds.   The {\em maximum} mass-loss rate a given CR energy flux can sustain is significantly larger than the galaxies' star formation rate (eq. \ref{eq:mdotmaxvsmdotstar}).    However, this is only realizable for relatively small diffusion coefficients, $\kappa \lesssim r_0 c_i$.  If, on the other hand, $\kappa \gg r_0 c_i$, $\dot M_w/\dot M_* \lesssim 1$ (eq. \ref{eq:mdotf}) and thus less dynamically important.   In \S \ref{sec:pion} we estimate the diffusion coefficient in star-forming galaxies using gamma-ray constraints on pion losses.

\section{Numerical Simulations}
\label{sec:numerics}
\subsection{Equations}
\label{sec:methods}

We solve the time-dependent cosmic ray hydrodynamic equations based on the two moment approach as developed by \cite{Jiang2018} in one dimensional spherical polar coordinates.\footnote{The two-moment approach enables a much more accurate treatment of CR streaming, which is not essential for the present paper but will be in later publications in this series.}  This algorithm has been implemented in the magneto-hydrodynamic code {\sf Athena++} \citep{Stone2020}.   As in \S \ref{section:analytic}, we use an isothermal equation of state  for the gas with isothermal sound speed $c_i$ and take the gravitational potential to be $\Phi=2V_g^2\ln r$.

Neglecting CR streaming, the full set of equations for gas density $\rho$, flow velocity $v$, comic ray energy density $E_c$ and flux $F_c$ in 1D spherical polar coordinates are
\begin{eqnarray}
\frac{\partial \rho}{\partial t}+\frac{1}{r^2}\frac{\partial}{\partial r}\left(r^2\rho v\right)&=&0,\nonumber\\
\frac{\partial \left(\rho v\right)}{\partial t}
+\frac{1}{r^2}\frac{\partial }{\partial r}\left(r^2 \rho v^2\right)&=&-\rho\frac{\partial \Phi}{\partial r}-c_i^2\frac{\partial \rho}{\partial r} +\sigma_c\left[F_c-v(E_c+p_c)\right],\nonumber\\
\frac{\partial E_c}{\partial t}+\frac{1}{r^2}\frac{\partial\left( r^2F_c\right)}{\partial r}&=&
- v \, \sigma_c [F_c-v(E_c+p_c)],\nonumber\\
\frac{1}{V_m^2}\frac{\partial F_c}{\partial t}+\frac{\partial p_c}{\partial r}&=&-\sigma_c\left[F_c-v(E_c+p_c)\right].
\label{eq:CR2mom}
\end{eqnarray}
Here the cosmic ray pressure is $p_c=E_c/3$.  The reduced speed of light is $V_m$, which is chosen to be much larger than $v$ in the whole simulation box.  For CR transport by spatially independent diffusion, $\sigma_c$ is a constant and is related to the  diffusion coefficient used elsewhere in this paper by $\kappa\equiv 1/(3 \sigma_c)$.\footnote{\citet{Jiang2018} follow the conventions of the radiation transfer literature in which the diffusive flux is $F_c = -\sigma_c^{-1} \nabla p_c$. In the rest of this paper, we follow the conventions of the CR literature and define the diffusive flux as $F_c = -\kappa \nabla E_c$ (eq. \ref{cr_energy}).  This accounts for the factor of 3 relating $\kappa$ and $\sigma_c^{-1}$.}

Substituting the right-hand-side of the fourth of equations \ref{eq:CR2mom} into the 3rd term on the right-hand-side of the 2nd of equations \ref{eq:CR2mom} produces the usual $\partial p_c/\partial r$ cosmic-ray pressure gradient in the momentum equation, along with another term related to the time variation of the CR flux.  Likewise, the term on the right-hand-side of the 3rd of equations \ref{eq:CR2mom} becomes the usual $v \partial p_c/\partial r$ term related to CR pdV work, again with another term related to the time variation of the CR flux.  The steady state versions of equations \ref{eq:CR2mom} are thus equivalent to the steady state equations \ref{eq:mdot}-\ref{eq:pc} solved in \S \ref{section:analytic}.  

\begin{table}
	\caption{Summary of Units for Numerical Simulations. $c_i$ is the gas isothermal sound speed (assumed to be a constant), $\rho_0$ is the gas density at the base of the wind (at radius $r = r_0$), and $p_{c,0}$ is the base CR pressure.}
\centerline{	\begin{tabular}{c|cc}
Quantity & Symbol & Units \\ \hline
Radial Velocity & $v$ & $c_i$ \\
`Isothermal' CR Sound Speed & $c_c \equiv \left(\frac{p_c}{\rho}\right)^{1/2}$ & $c_i$ \\
Gravitational Velocity  & $V_g$ & $c_i$ \\
Density & $\rho$ & $\rho_0$ \\
CR Pressure & $p_c$ & $p_{c,0}$ \\
CR Flux & $F_c$ & $c_i p_{c,0}$ \\
CR Diffusion Coefficient & $\kappa$ & $c_i r_0$ \\
\hline
\label{tab:units}
\end{tabular}}
\end{table}

\subsection{Initial and Boundary Conditions}
For each simulation, we pick gas density $\rho_0$ and cosmic ray pressure $p_{c,0}$ at the bottom boundary $r_0$ and then initialize the gas density and cosmic ray energy density at each radius $r$ as
\begin{eqnarray}
\rho(r)&=&\rho_0 \left(r/r_0\right)^{-V_g^2/c_i^2},\nonumber\\
E_c(r)&=&3p_{c,0}\left(r/r_0\right)^{-V_g^2/c_i^2}.
\end{eqnarray}
We apply floor values of $10^{-4}$ for $\rho/\rho_0$ and $p_c(r)/p_{c,0}$ in the initial condition. The flow velocity and cosmic ray flux are set to be zero in the whole simulation domain initially.

For the bottom boundary condition at $r_0$, the cosmic ray energy density $E_c$ and gas density $\rho$ are fixed to be the initial values. The flow velocity in the ghost zones is set such that mass flux $\rho v r^2$ is continuous from the last active zone to the ghost zones. We keep the gradient of the cosmic ray flux  continuous across the bottom boundary. For the boundary condition at the top of the simulation domain, we keep the gradient of $\rho, E_c$ and $F_c$  continuous across the boundary. The flow velocity at the top boundary is  set by requiring the mass flux $\rho v r^2$ to be continuous.

\subsection{Overview of Simulations}

Table \ref{tab:compare} summarizes our suite of simulations.   The key physical parameters of each simulation are $V_g/c_i$ and $p_{c,0}/\rho_0 c_i^2$, as in the analytics of \S \ref{section:analytic}, as well as the diffusion coefficient $\kappa$ (in units of $r_0 c_i$).   The key numerical parameters are the resolution, reduced speed of light, and box size.   

The units for the results of our numerical simulations are summarized in Table \ref{tab:units}: gas density is in units of the base density $\rho_0$, speeds are in units of the gas isothermal sound speed $c_i$, CR pressure is in units of the base CR pressure $p_{c,0}$, and CR fluxes are in units of $p_{c,0} c_i$.

\begin{figure}
\centering
\includegraphics[width=82mm]{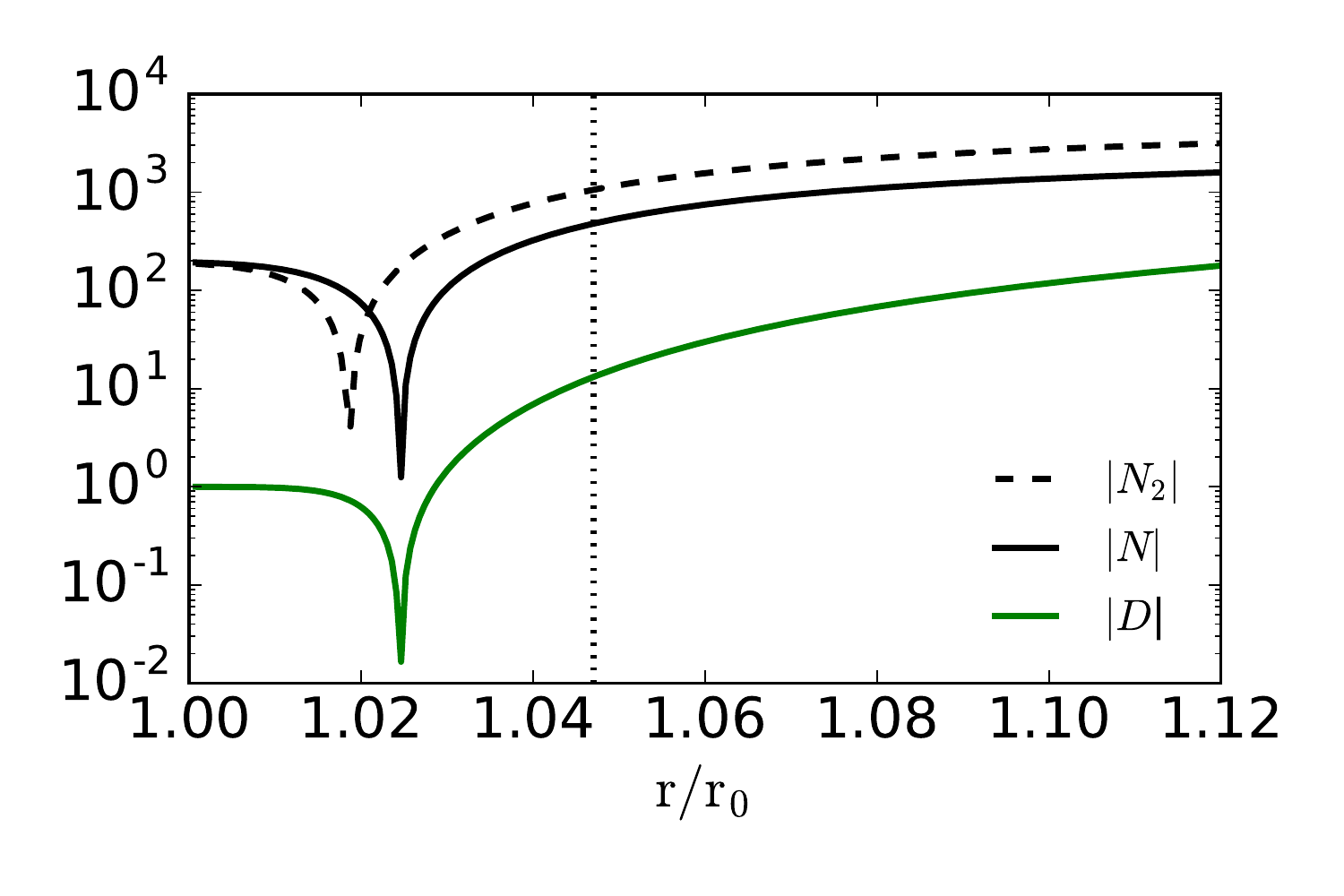}
\vspace{-0.2cm}
\caption{Numerator and denominator (eqs \ref{eq:ND}) of the steady state wind equation evaluated for our $\kappa=10$, $V_g=10$ simulation.  $N_2$ is the approximation to the numerator of the wind equation used in our analytics (eq. \& \ref{eq:N2}). The vertical dotted line is the analytic estimate of the location of the critical point (eq. \ref{eq:rc}).  The numerical simulation reaches a steady state that passes through a critical point close to that predicted by our (approximate) analytics.}
\label{fig:sonic_diff}
\end{figure}

\begin{figure}
\centering
\includegraphics[width=82mm]{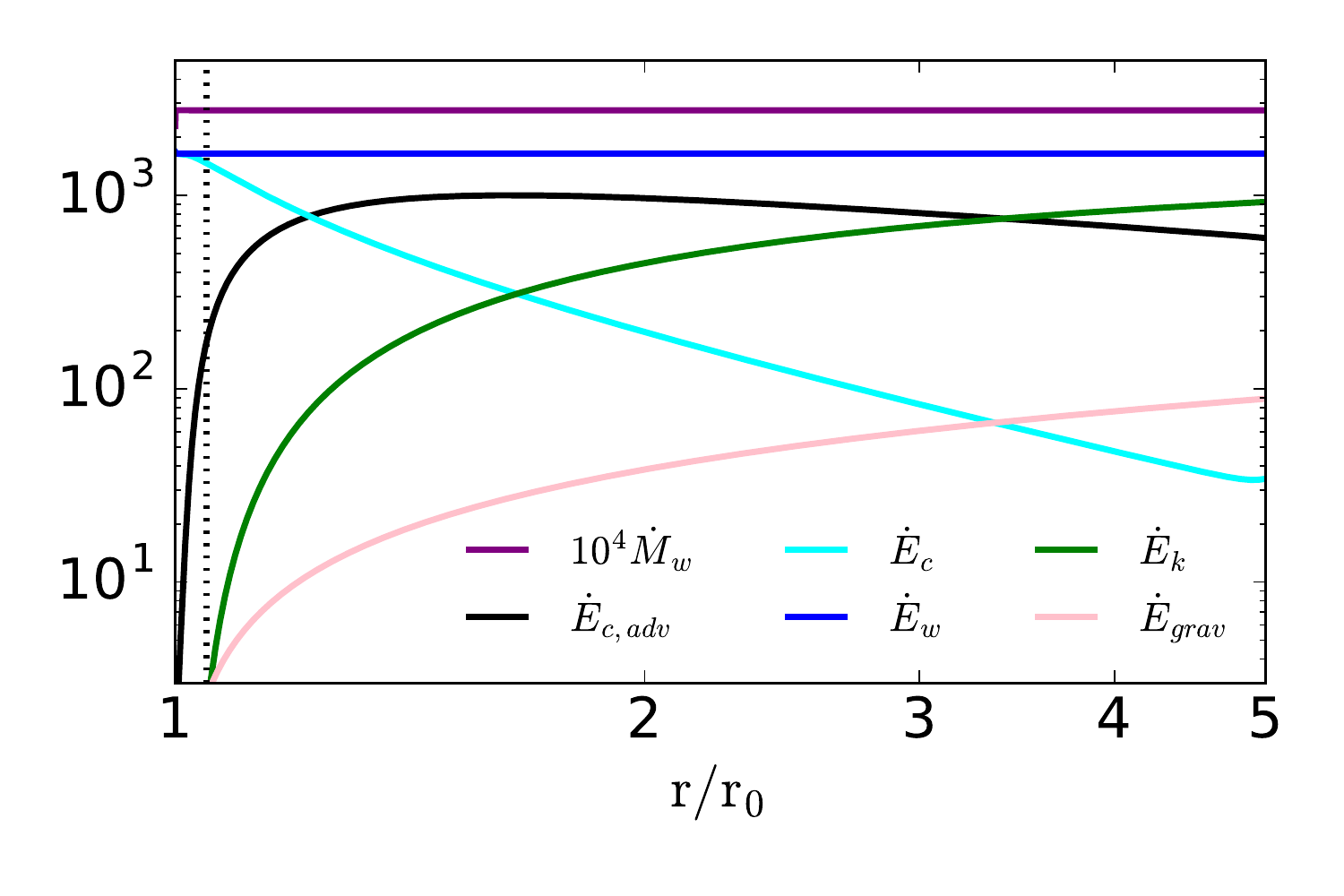}
\vspace{-0.2cm}
\caption{Energetics of the wind evaluated for our $\kappa=10$, $V_g=10$ simulation.   The total energy flux ($\dot E_w$; see eq.~\ref{eq:Edotdiff}) and mass flux ($\dot M_w$) are independent of radius.  At the base of the wind, nearly all of the energy is carried by the diffusing CRs ($\dot E_c$). Advection of CR enthalpy by the motion of the gas ($\dot E_{c,adv}$) is important at intermediate radii while at large radii most of the energy is in the kinetic energy of the outflowing gas ($\dot E_{k}$).  For the logarithmic potential used in our simulations, we define $\dot E_{grav}$ using $\Phi=2 V_g^2 \ln(r/r_0)$ so that $\dot E_{grav} > 0$.}
\label{fig:Edot_diff}
\end{figure}

Nearly all of our numerical solutions reach a steady state state that does not evolve significantly in time once we run for a few box crossing times.   Because the solutions reach a steady state, they are independent of the reduced speed of light $V_m$, which only enters into the time-dependent terms in equations \ref{eq:CR2mom}.

\begin{figure*}
\centering
\includegraphics[width=174mm]{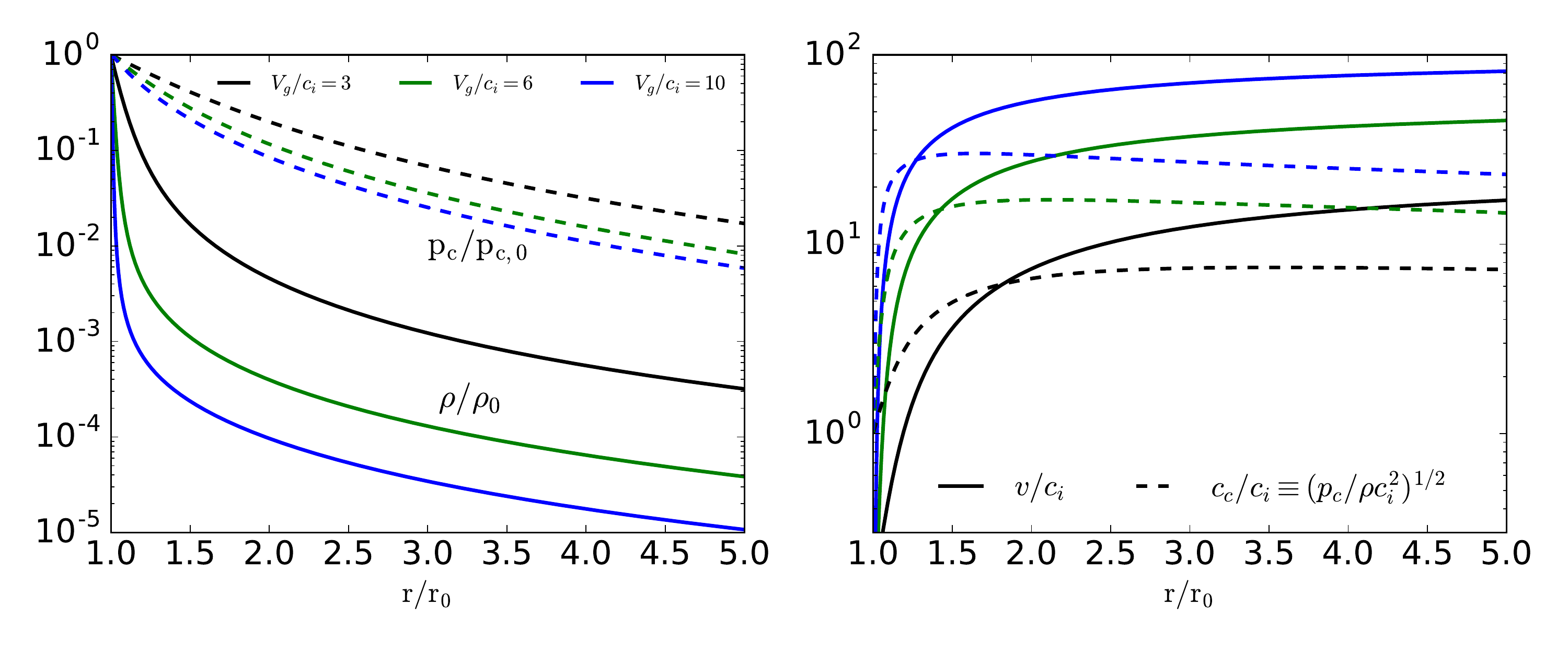}
\vspace{-0.2cm}
\caption{Dependence of density and pressure profiles ({\em Left}) and velocity and CR sound speed profiles ({\em Right}) on the depth of the gravitational potential $V_g$ (in units of $c_i$, the isothermal gas sound speed).   All solutions are for $\kappa=10$ and equal gas and CR pressures at the base of the wind ($p_{c,0}=\rho_0 c_i^2$).  The CR pressure profiles are relatively independent of the gravitational potential, being largely set by rapid CR diffusion.   By contrast, the density falls off more gradually, and the flow accelerates more gradually, for weaker gravity.}
\label{fig:Vg}
\end{figure*}

\citet{Drury1986} showed that rapid CR diffusion drives sound waves unstable in the presence of a background CR pressure gradient, so it is not a priori obvious that our wind solutions should reach a steady state.  We show in Appendix \ref{sec:appendixA} that the growth rate of this instability is too slow to grow significantly in galactic winds, unless the gas sound speed is extremely small (the $V_g=200$ simulation, which is our lowest $c_i$ simulation, is the only one to show significant time dependence; see Fig. \ref{fig:app}).  We also derive the linear WKB dispersion relation for sound waves and entropy modes in the {\em two-moment} CR system of equations, and show that the linear modes are stable, consistent with the simulations.   

\subsection{Numerical Solutions For $\kappa \gtrsim r_0 c_i$}

Figure \ref{fig:flow_diff} shows the resulting density, CR pressure, fluid velocity,  and CR sound speed profiles for $V_g/c_i = 10$, varying $\kappa$ from $0.33-30$ (in units of $r_0 c_i$; Table \ref{tab:units}).     The vertical dotted line in Fig \ref{fig:flow_diff} (as well as Figs. \ref{fig:sonic_diff} \& \ref{fig:Edot_diff} below) is the critical point predicted in equation \ref{eq:rc}, namely $r_s \simeq 1.047 r_0$.

The left panel of Figure \ref{fig:flow_diff} shows that for all of the solutions, the CR pressure falls off much more slowly than the density near the base.   This is because the gas density scale-height is initially quite small, $\sim r_0 (c_i/V_g)^2 \sim 10^{-2} r_0$ while the CR pressure scale-height is significantly larger due to CR diffusion $\sim h_c r_0 \sim 0.25 r_0$ (eq. \ref{eq:hc}). For all of the solutions, the hydrostatic equilibrium approximation for the density profile (eq. \ref{eq:rho}) is reasonable at very small radii.   There is, however, a bifurcation in the density profiles as a function of $\kappa$.   The solutions with $\kappa$ = 0.33 and 1 are closer to hydrostatic over the entire radial range while for $\kappa \gtrsim 1$ the density falls off much more rapidly at large radii.   The right panel of Figure \ref{fig:flow_diff} shows that this bifurcation in density profiles corresponds to a bifurcation in velocity profiles.   The solutions with $\kappa \gtrsim 3$ drive strong winds that accelerate to a velocity larger than the escape speed by $r \sim 1.2 r_0$.   By contrast, for $\kappa =1$ the velocities are much lower and for $\kappa=0.33$ the solution is subsonic for all of the radii shown in Figure \ref{fig:flow_diff}.   We return to the lowest $\kappa$ simulation in \S \ref{sec:lowkapnum} and show that it does produce a wind, but one whose character is very different from the higher $\kappa$ simulations focused on here.

Figure \ref{fig:sonic_diff} quantifies the extent to which the numerical solution satisfies the time steady critical point conditions, for the $\kappa = 10$ numerical simulation.    Specifically, from equation \ref{eq:wind1} we define the denominator and numerator of the critical point equation to be 
\be
D \equiv v^2 - c_i^2 \ \ \  {\rm and} \ \ \ N \equiv \frac{2 c_i^2}{r} - \frac{1}{\rho}\frac{dp_c}{dr} - \frac{2 V_g^2}{r}
\label{eq:ND}
\ee
and we further define 
\be 
N _2\equiv \frac{2 c_i^2}{r} + \frac{p_{c,0} r_0}{\rho r^2 h_c} - \frac{2 V_g^2}{r}
\label{eq:N2}
\ee
to be the numerator of the critical point equation under the approximation that $\dot E_c$ is constant as a function of radius, which was used in our analytic derivations in \S \ref{section:analytic}.  Figure \ref{fig:sonic_diff} shows that $N$ and $D$ pass through zero at the same point, as expected for a time steady solution that passes through the critical point.    The location of the critical point is close to the prediction in equation \ref{eq:rc}, which is shown by the vertical dotted line in Fig. \ref{fig:sonic_diff}; the deviation is because the analytics makes the approximation that the hydrostatic density profile (eq. \ref{eq:rho}) is valid all the way to the sonic point.   Equation \ref{eq:N2} (used in our analytics) is also a reasonable, though not perfect, approximation to the true numerator in the critical point equation.

Figure \ref{fig:Edot_diff} quantifies the radial profiles of the mass-loss rate $\dot M$ and various contributions to the wind energy flux in the numerical solution for $\kappa = 10$.   The mass-loss rate $\dot M$ is independent of radius as expected for a steady state solution.      The contributions to the energy flux shown in Figure \ref{fig:Edot_diff} are defined in and below equation \ref{eq:Edotdiff} (we do not show the gas enthalpy/advective flux, which is negligible).   We note that for the logarithmic potential used here, we define $\Phi=2 V_g^2 \ln(r/r_0)$.  The zero point of the potential is thus at the base of the wind ($r=r_0$), rather than at infinity, as is more common; the latter is not possible given the divergence of the logarithmic potential as $r \rightarrow \infty$.  As a result of this choice of zero point, $\dot E_{grav} > 0$.

Figure \ref{fig:Edot_diff} shows that interior to the critical point, the energy flux is dominated by CR diffusion, as assumed in the analytic calculation in \S \ref{section:analytic}. At intermediate radii the CR advective flux dominates the energy transport while eventually at large radii, when the CR pressure has accelerated the gas to well above the CR sound speed and escape speeds, the gas kinetic energy flux dominates.  Note that the gravitational contribution to the energy flux is small at all radii, consistent with $\dot M_w \ll \dot M_{\rm max}$ and $v > V_g$. The qualitative features of Figure \ref{fig:Edot_diff} are the same for $\kappa = 3-30$.  For $\kappa=1$, however, most of the energy at the outer edge of the domain is in CR enthalpy rather than gas kinetic energy.\footnote{This implies that $v(r_{out})=0.5 V_g = 5 c_i$ in Table \ref{tab:compare} is a lower limit.   If all of the CR enthalpy is converted to kinetic energy, the final velocity would be $\simeq 1.8 V_g$, close to the analytic estimate in equation \ref{eq:vinfdiff}.}   For $\kappa = 0.33$, the energetics of the flow is yet more different from that in Figure \ref{fig:Edot_diff}, as we discuss in \S \ref{sec:lowkapnum}.

Figure \ref{fig:Vg} shows how the $\kappa = 10$ solutions depend on the depth of the gravitational potential $V_g$ (in units of the isothermal sound speed $c_i$).   The solutions are qualitatively very similar.   The primary difference is that weaker gravity (lower $V_g$) implies a larger density scale-height at the base of the wind.  Correspondingly, the wind accelerates significantly more slowly.

\subsection{Low $\kappa \lesssim r_0 c_i$ Solutions}
\label{sec:lowkapnum}

The analytic derivations in \S \ref{section:analytic} primarily assumed $\kappa \gtrsim r_0 c_i$.  Our numerical simulations with $\kappa \gtrsim r_0 c_i$ are consistent with the properties of these analytic solutions, as we discuss further in the next section.  Here we present numerical wind models with $\kappa \lesssim r_0 c_i$, which differ dramatically from their high $\kappa$ counterparts.  We stress that the higher $\kappa$ solutions are likely the most astrophysically relevant, given estimates of diffusion coefficients from MW CR data (\S \ref{section:analytic}) and gamma-ray data from pion decay (\S \ref{sec:pion}).

\begin{figure}
\centering
\includegraphics[width=82mm]{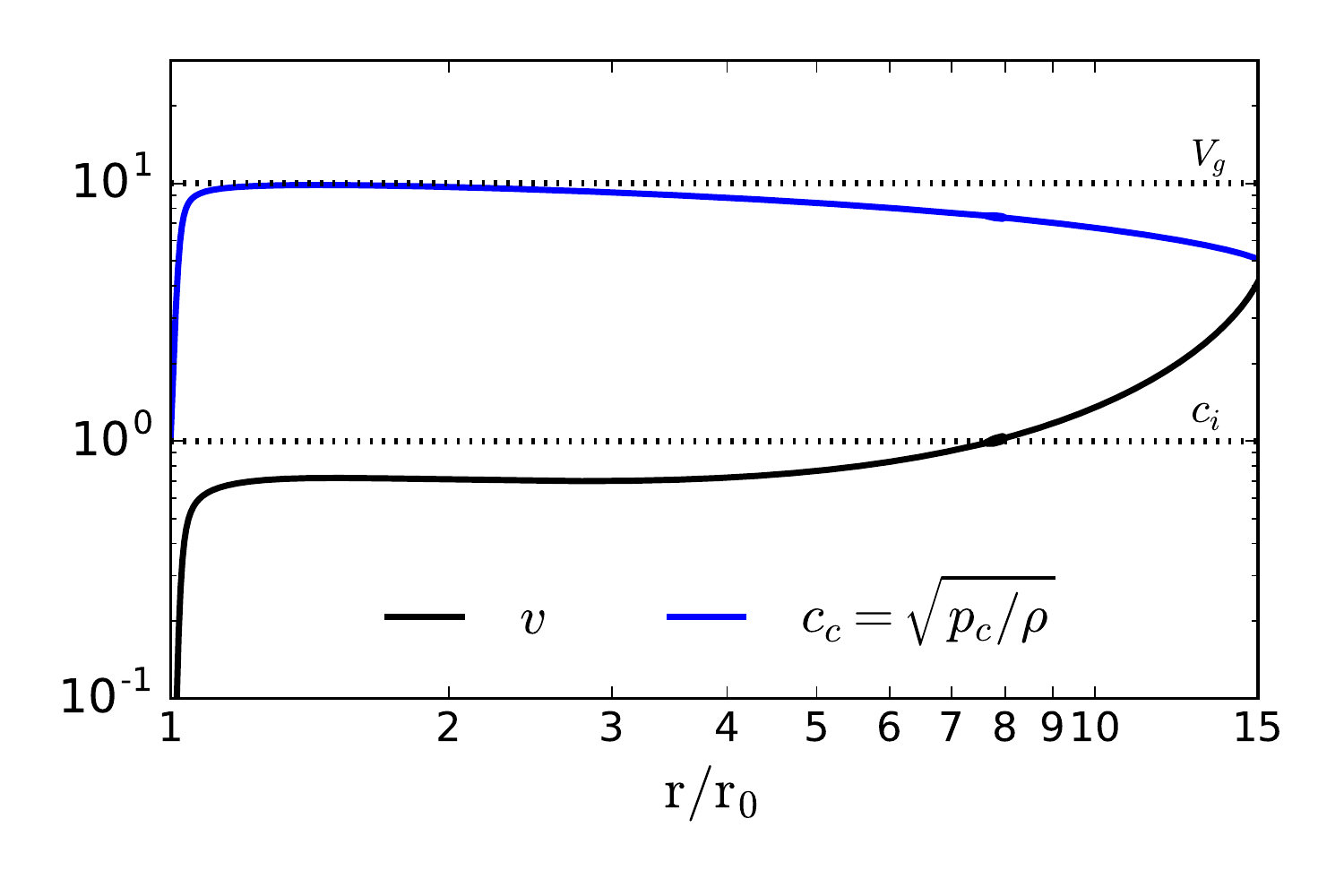}
\includegraphics[width=82mm]{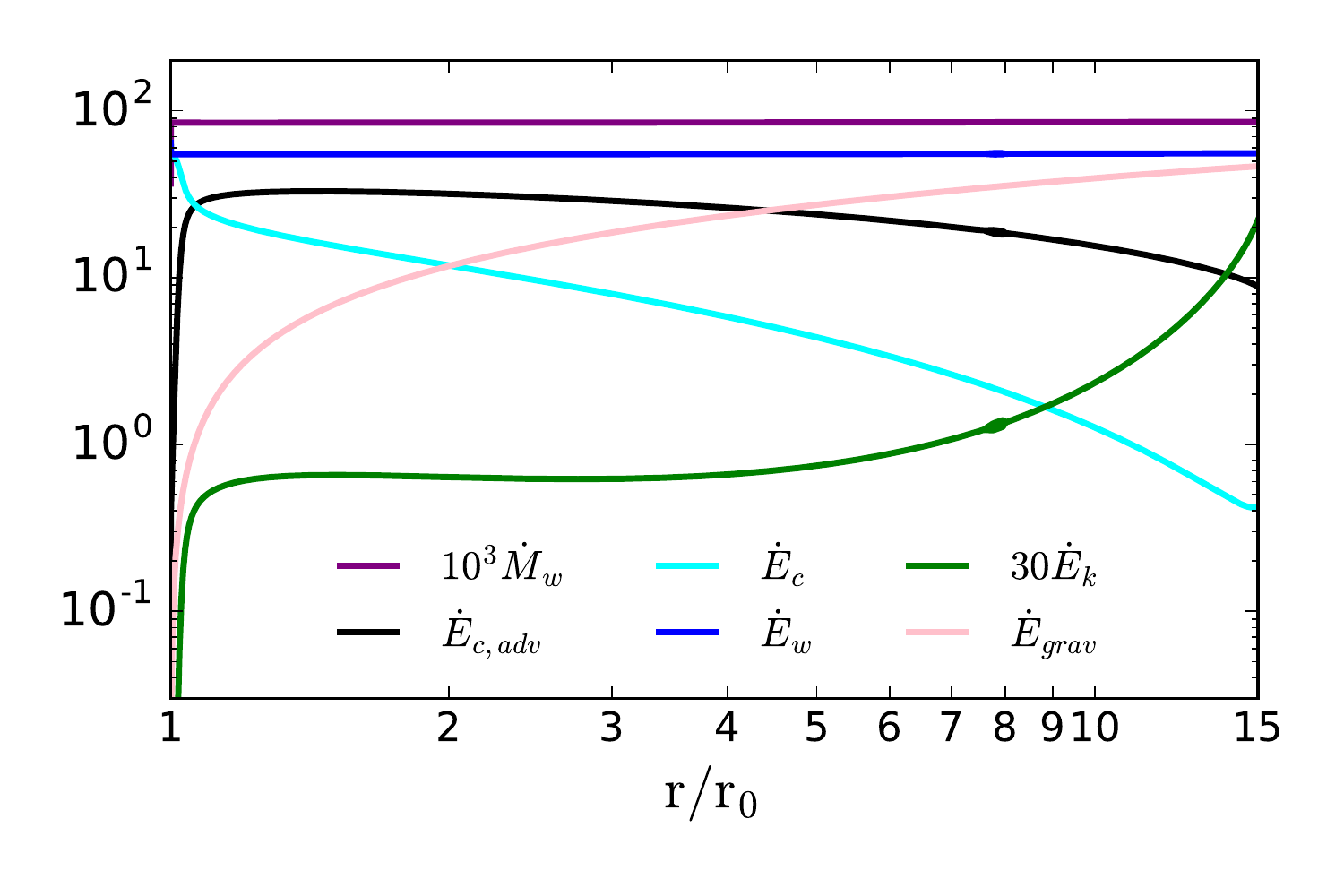}
\vspace{-0.2cm}
\caption{Kinematics (top) and energetics (bottom) of the wind for our $\kappa=0.33$, $V_g=10$ simulation, which differs dramatically from the higher $\kappa$ simulations.  The flow accelerates very gradually, reaching the sonic point ($v/c_i=1$) only at $r \sim 8 r_0$, compared to $r \simeq 1.02 r_0$ for high $\kappa$ solutions.  The flow velocity is also significantly lower ($\lesssim V_g$) than for the high $\kappa$ solutions (Fig. \ref{fig:flow_diff}).  At the base of the wind, nearly all of the energy is carried by the diffusing CRs ($\dot E_c$).   However, almost all of this energy is used to overcome the gravitational  potential, leading to $\dot E_{grav} \simeq \dot E_w$ at large radii (note that $\dot E_{grav} > 0$ for our logarithmic potential with $\Phi = 2 V_g^2 \ln(r/r_0)$).    This corresponds to $\dot M_w \simeq \dot M_{\rm max}$ (eq. \ref{eq:mdotmax}) and is a CR analogue of photon-tired winds in stellar wind theory \citep{Owocki1997}.   For larger $\kappa$ the energy lost to gravity is negligible and nearly all of the CR energy at the base goes into CR enthalpy and kinetic energy at larger radii (see Fig. \ref{fig:Edot_diff}).}
\label{fig:Edot_diff_lowkap}
\end{figure}

Figure \ref{fig:Edot_diff_lowkap} shows the kinematics (top) and energetics (bottom) of the wind for our $\kappa=0.33$, $V_g=10$ simulation, which differs significantly from the higher $\kappa$ simulations. The velocity profile in Figure \ref{fig:Edot_diff_lowkap} is shown over a larger range of radii than was plotted in Figure \ref{fig:flow_diff}.   The flow accelerates very gradually reaching the sonic point only at $r_s \sim 8 r_0$\footnote{Note that there are small oscillations in the solution near the sonic point; these remain at late times even once the solution settles into a steady state.  We do not have a definitive explanation for these oscillations but they are not present in any of our $\kappa \gtrsim r_0 c_i$ solutions (the majority in Table \ref{tab:compare}).}, compared to $r_s \simeq r_0$ for the high $\kappa$ solutions.  The most striking property of the energetics is that $\dot E_{grav} \simeq \dot E_w$ (see eq.~\ref{eq:Edotdiff}) at large radii while at the base of the wind, nearly all of the energy is carried by CR diffusion ($\dot E_c$); recall that for our logarithmic potential $\Phi = 2 V_g^2 \ln(r/r_0)$ and so $\dot E_{grav} > 0$.    The fact that $\dot E_{grav}(r_{out}) \simeq \dot E_c(r_0)$ implies that most of the energy supplied to CRs at the base goes into lifting (almost hydrostatically) the gas out to large radii.   As a result, the gas kinetic energy is a minor contribution to the energy flux at large radii, in marked contrast to the high $\kappa$ simulation shown in Figure \ref{fig:Edot_diff}.    The condition $\dot E_{c} \simeq \dot E_{grav}$ is precisely the condition for the CR-analogue of photon-tired winds; this is also the regime in which $\dot M_w \sim \dot M_{\rm max}$ (eq. \ref{eq:mdotmax}).   For our simulations, we can define the escape velocity from the base of the wind to be the velocity needed to reach the outer edge of the box, i.e, $v_{\rm esc}(r_0) = 2 V_g \sqrt{\ln(r_{out}/r_0)}$.  Equation \ref{eq:mdotmax} then becomes
\be
\frac{\dot M_{max}}{\dot M_0} \simeq \frac{3}{2 h_c \ln(r_{\rm out}/r_0)} \frac{\kappa}{r_0 c_i} \frac{c_{c,0}^2}{V_g^2}
\label{eq:mdotmaxsim}
\ee 
Table 1 gives the simulation mass-loss rates in units of $\dot M_{max}$.  For $\kappa = 0.33 r_0 c_i$, $\dot M_w \simeq 0.78 \dot M_{max}$, close to the maximum value allowed by energy conservation. This is consistent with the energetics in Figure \ref{fig:Edot_diff_lowkap}.   For comparison, we note that $\dot M_w/\dot M_{\rm max} = 0.85, 0.78, 0.75, 0.42, 0.15, 0.054, 0.018$ for our simulations with $\kappa/r_0 c_i = 0.11, 0.33, 1, 2, 3.3, 10, \& 30$, respectively.  This is consistent with the expectation from equation \ref{eq:mdotwvsmdotmax} that $\dot M_w/\dot M_{max} \propto 1/\kappa$ for $\kappa \gtrsim r_0 c_i$.   The bifurcation in density and velocity profiles in Figure \ref{fig:flow_diff} at $\kappa \simeq r_0 c_i$ thus corresponds to solutions with $\dot M_w \sim \dot M_{max}$ ($\kappa \lesssim r_0 c_i$, nearly hydrostatic, slower acceleration) vs. those with $\dot M_w \ll \dot M_{max}$ ($\kappa \gtrsim r_0 c_i$, rapid acceleration).   More generally, the last column in Table \ref{tab:compare} shows that for solutions with $\kappa > r_0 c_i$ nearly all of the CR energy supplied at the base goes into kinetic energy or enthalpy of the wind at large radii.   By contrast, for solutions with $\kappa < r_0 c_i$, the asymptotic kinetic and enthalpy flux is small compared to the CR energy flux at the base because most of the energy is lost escaping the gravitational potential.   

The importance of gravity for the low $\kappa$ solution in Figure \ref{fig:Edot_diff_lowkap} implies that some of the details of the solution -- though not the fact that $\dot M_w \simeq \dot M_{max}$ -- will likely be sensitive to the form of the gravitational potential.   In particular, we suspect that the exact acceleration profile for the gas, including the final terminal speed (which depends on the small residual energy {\em not} lost to work against gravity), will depend on the details of the potential.   

\subsection{Synthesis of Analytics vs. Numerics}

\begin{figure}
\centering
\includegraphics[width=84mm]{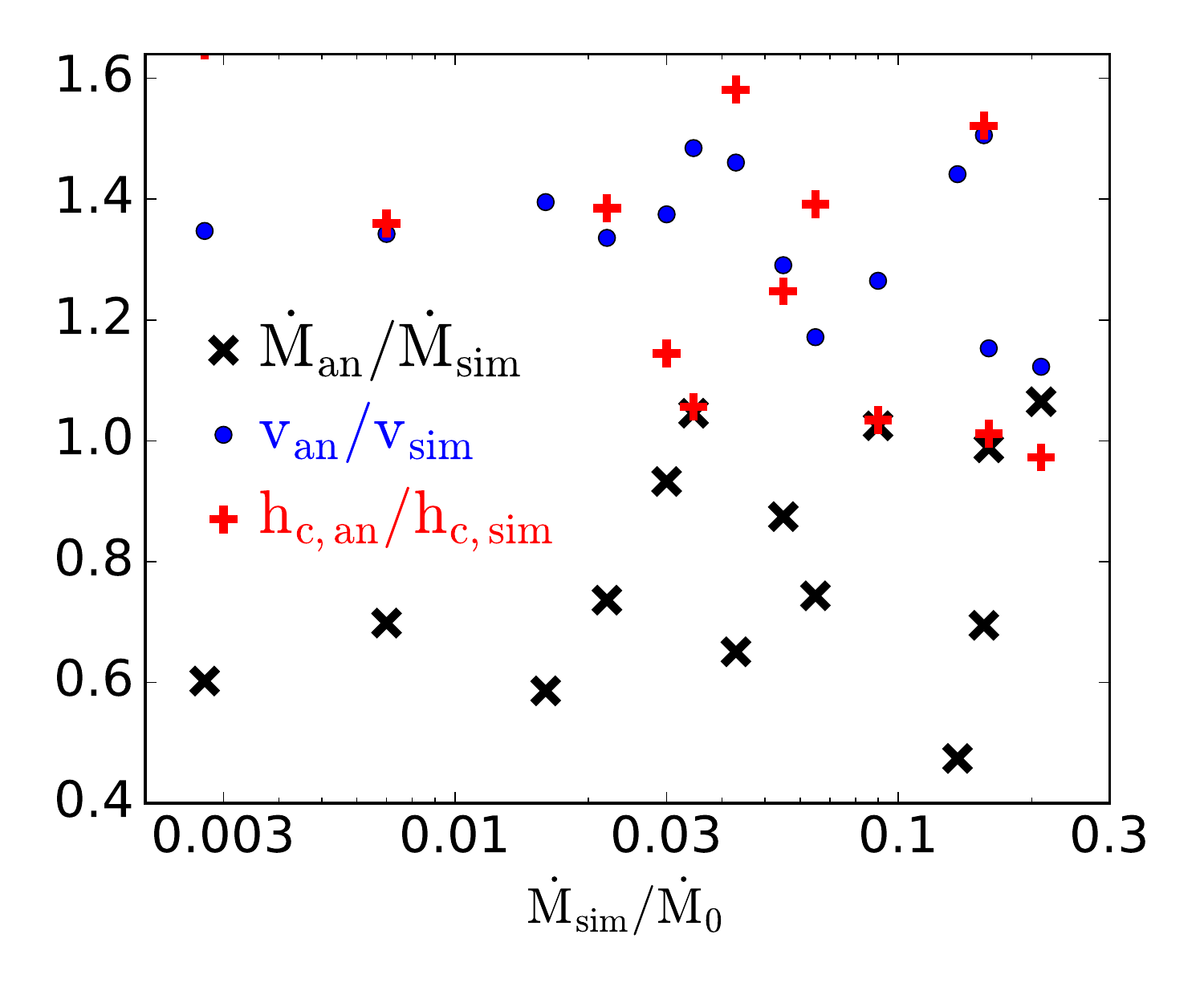} 
\caption{Ratio of the analytically predicted mass-loss rate (eq. \ref{eq:mdot3} with $h_c$ from eq. \ref{eq:hc}), terminal velocity (eq. \ref{eq:vinfdiff}), and base CR scale-height (eq. \ref{eq:hc}) compared to the simulation results.  The agreement is good for the full range of simulated parameters.  Multiplying $h_c$ in eq. \ref{eq:hc} by $\simeq 0.75$ would remove the small systematic offset in $h_c$ and $\dot M_w$.}
\label{fig:mdotfin}
\end{figure}

Figure \ref{fig:mdotfin} compares our  numerical and analytic solutions for the mass-loss rate, terminal velocity, and base CR scale-height.  We focus on the $\kappa > r_0 c_i$ solutions for which we have the most detailed analytic estimates.   For $\kappa < r_0 c_i$, $\dot M_{w} \simeq \dot M_{max}$ (eq. \ref{eq:mdotmaxsim}), but we do not have a prediction for the terminal velocity in this regime.

Overall, Figure \ref{fig:mdotfin} shows that the analytics and numerics agree well using our analytically estimated base CR scale-height (eq. \ref{eq:hc}). The latter in turn agrees well with the full numerical simulations (bottom panel of Fig. \ref{fig:mdotfin}). There is a slight systematic offset in the mean analytical estimates of $h_c$ and $\dot M$ that could be removed by multiplying equation \ref{eq:hc} by $\simeq 0.75$.

The agreement in Figure \ref{fig:mdotfin} holds over a factor of $\sim 30$ in CR diffusion coefficient, $\sim 100$ in base CR pressure, and $\sim 10^4$ in the ratio of gravity to gas pressure ($\propto V_g^2$) which is also a proxy for galaxy mass; see Table \ref{tab:compare} for the full range of simulations.  Figure \ref{fig:mdotfin} also compares the analytic estimate of the wind terminal velocity (eq. \ref{eq:vinfdiff}) to the speed in the simulations at the top of the box.  There is again reasonably good agreement, though the analytic speeds tend to be somewhat ($\sim 30 \%$) larger than the simulations.   This is primarily because some of the energy remains in the CRs in the simulations given the finite outer radius of the computational domain, while the analytic estimate assumes that all of the CR energy has been transferred to the gas.

\section{Models Calibrated to Gamma-ray Observations}

\label{sec:pion}

The analytic and numerical calculations in \S \ref{section:analytic} and \ref{sec:numerics} show how the properties of galactic winds driven by CR diffusion depend on the diffusion coefficient.  For example, winds lose most of their energy to gravity if $\kappa \lesssim r_0 c_i$ (Table \ref{tab:compare}), while if $\kappa \gtrsim r_0 c_i$, the terminal speed of the wind $\propto \kappa^{1/2}$ (eq. \ref{eq:vinfdiff}) and the base CR pressure and the wind mass-loss rate  depend on $\kappa$ for a given star formation rate (eq. \ref{eq:mdotedot}).  To assess the implications of our results for CR driven galactic winds we must thus have a handle on $\kappa$ in other galaxies.  Unfortunately, however, there is still sufficient uncertainty in CR transport that it is non-trivial to determine from first principles if the transport is indeed diffusive (vs. streaming), let alone the value of the diffusion coefficient in different phases of the ISM and for the full range of conditions realized in galaxy formation.   As a result, we appeal instead to observations to calibrate physically reasonable diffusion coefficients (see also \citealt{Chan2019, Hopkins2020a}) and then use those, together with the results of \S  \ref{section:analytic} and \ref{sec:numerics}, to quantify the implications for CR-driven galactic wind models.

The non-thermal radio and gamma-ray emission from galaxies provide direct observational constraints on the properties of CRs in external galaxies.  These in turn inform the CR pressure and diffusion coefficient that are critical for setting the strength of CR-driven galactic winds.   In this section, we provide simple estimates to elucidate these constraints.   We focus on the gamma-ray emission from pion decay in star forming galaxies observed by Fermi \citep{Fermi2010,Fermi2010b,Fermi2012}, rather than non-thermal radio emission, despite the fact that there are far more observations available for the latter (see \citealt{Lacki2010}).   The reason is that the Fermi data directly constrains the properties of the GeV protons that dominate the CR energy density.

We consider a simple one-zone model in which a galaxy is characterized by its size $r_0$, gas surface density $\Sigma_g$, and isothermal/turbulent velocity $c_i$.  The cosmic-ray scale-height is $H_c$  while the gas scale-height is $H_g$.  For CR diffusion we expect $H_c > H_g$, as is indeed the case in our simulations in \S \ref{sec:numerics}.     The gamma-ray luminosity is set by the ratio of the timescale for pion losses $t_\pi \simeq t_0 \, (n_\pi/n_{eff})$ (where  $t_0=5 \times 10^7$ yrs and $n_\pi=1$ cm$^{-3}$ are set by the cross-section for pion production and the effective density $n_{eff}$ is defined below) to the timescale for CRs to escape by diffusion, $t_{\rm diff} \simeq H_c^2/\kappa$.  The effective density $n_{eff}$ is the density averaged over the region the CRs are diffusing through and so is given by $\simeq n H_g/H_c$ where $n$ is the mid-plane density of the galaxy.   For reasons that will become clear, we choose to express $t_\pi$ in terms of gas surface density by using $\Sigma_g \simeq 2 \rho H_g$.    The ratio of the diffusion timescale to the pion loss timescale is then
\be
\frac{t_{\rm diff}}{t_{\pi}} \simeq \frac{\Sigma_g H_c}{2 \kappa \rho_\pi t_0 } 
\label{eq:tpi}
\ee
where $\rho_\pi=n_\pi m_p \simeq 1.67 \times 10^{-24}$ cm$^{-3}$. Equation \ref{eq:tpi} shows that for a fixed ratio of hadronic losses ($t_\pi$) to CR escape ($t_{\rm diff}$) there is a degeneracy between the CR scale-height and diffusion coefficient, with $\kappa \propto H_c$.   This is consistent with the known degeneracy between CR diffusion coefficient and `halo size' in the literature (e.g., Fig. 3 of \citealt{Trotta2011}).

The gamma-ray emission from pion decay in a star-forming galaxy is given by $L_\gamma \simeq 1/3 \dot E_\pi$ where $\dot E_\pi \simeq \dot E_c \min(t_{\rm diff}/t_\pi, 1)$ is the rate of energy loss to pion decay and the factor of $1/3$ quantifies the fraction of pion decay in neutral vs. charged pions.  Defining $\dot E_c = E_{cr} \dot M_*/m_*$ with\footnote{Note that $\epsilon_c$ defined in equation \ref{eq:edotc} is given by $\epsilon_c = E_{cr}/m_* c^2$.   In this section it is convenient to separate out $\epsilon_c$ into a part that depends on the initial mass function ($m_*$) and a part that depends on the CR energy supplied per SN ($E_{cr}$).}  $E_{cr} = 10^{50} E_{cr,50}$ the energy per SN going into CRs and $m_* \simeq 100 M_\odot$ the total stellar mass formed per core-collapse SNe, we find \citep{Lacki2011}
\be
L_\gamma  \simeq A_\gamma \, L_* \min\left(\frac{t_{\rm diff}}{t_\pi},1\right)
\label{eq:Lgam}
\ee
where $L_* \simeq \epsilon_{*} \dot M_* c^2$ is the total luminosity produced by star formation.  The factor $A_\gamma \simeq 3.3 \times 10^{-4} E_{cr,50}/(\epsilon_{*} m_* 17 M_\odot^{-1})$ quantifies the maximum possible gamma-ray luminosity from pion decay, which is realized in the limit $t_\pi < t_{\rm diff}$ when all of the CR proton energy is lost to pion decay before the CRs can escape \citep{Thompson2007,Lacki2011}.  Note that the factor $\epsilon_{*} m_* 17 M_\odot^{-1} \simeq 1$ is relatively independent of the stellar initial mass function because massive stars set the SN rate and the luminosity of a star-forming population.   To good accuracy, we can thus take $A_\gamma \simeq 3.3 \times 10^{-4} E_{cr,50}$.

\begin{figure*}
\centering
\includegraphics[width=174mm]{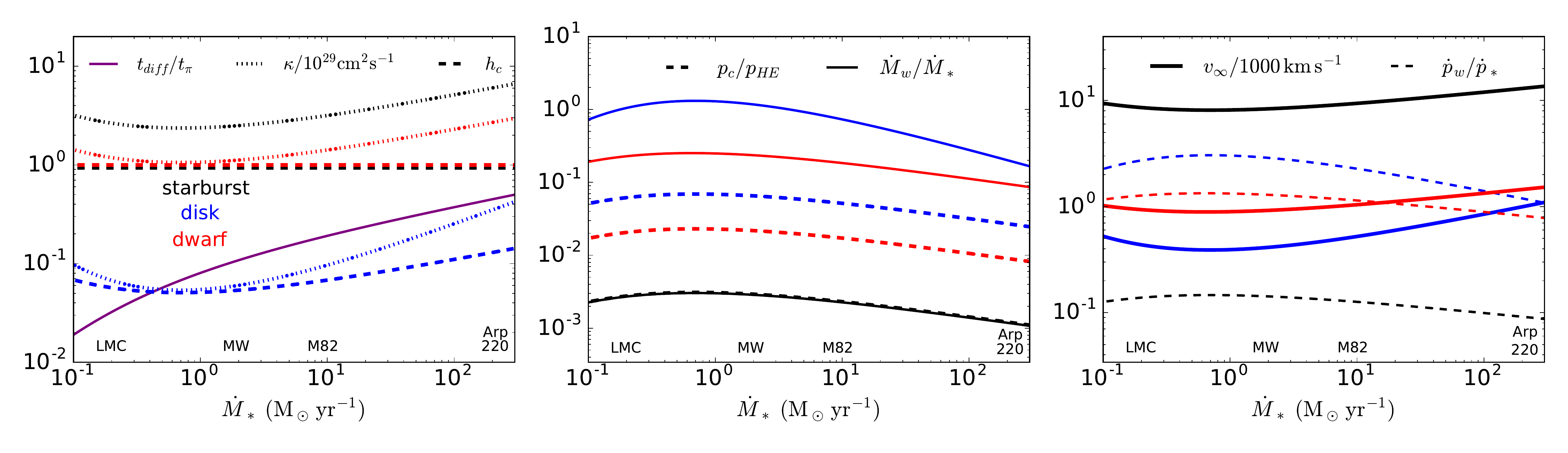}
\vspace{-0.4cm}
\caption{Empirically constrained cosmic-ray properties in star-forming galaxies inferred from gamma-ray observations of pion decay by Fermi, as a function of galaxy star formation rate.  All calculations assume gas isothermal sound speed $c_i=10 \kms$; scalings to other values of $c_i$ are shown in Figure \ref{fig:gammaci} and discussed in the text.   Star formation rates of example galaxies are indicated near the x-axis.  Three examples are considered corresponding to nuclear starbursts (black lines; $r_0=$ 0.3 kpc, $V_g=150 \kms$), dwarf galaxies (red lines; $r_0=1$ kpc, $V_g= 50\kms$), and star-forming disc galaxies (blue lines; $r_0=3$ kpc, $V_g=150 \kms$). {\em Left:}
$t_{\rm diff}/t_{\pi} \propto L_\gamma/L_*$ is the ratio of the CR diffusion time to the pion loss time and sets the gamma-ray luminosity of the galaxy (eq. \ref{eq:Lgam}).  We infer $t_{\rm diff}/t_{\pi}$ for different galaxy star formation rates using the Fermi correlation between $L_\gamma$ and star formation rate.
We then calculate the CR diffusion coefficient  $\kappa$ and the dimensionless CR scale-height $h_c = H_c/r_0$ using eq. \ref{eq:tpi}, eq. \ref{eq:SFR} and eq. \ref{eq:hc}.   {\em Middle:}  The fractional contribution of CR pressure to the pressure in the galactic disc ($p_{HE}=\pi G \Sigma_g^2 \phi)$ depends on the assumed size of the star-forming disc $r_0$, with nuclear starbursts having suppressed $p_c/p_{HE}$ because of rapid pion losses and CR diffusion in the dense nuclear regions.  The predicted mass-loss rate relative to the star formation rate $\dot M_w/\dot M_*$ is largest for the disc galaxy model.  {\em Right:}   Asymptotic wind speed $v_\infty$ and momentum flux in the wind $\dot p_w$ relative to the stellar radiation field ($\dot p_* = L_*/c$)}
\label{fig:gamma}
\end{figure*}

Observations of star-forming galaxies by Fermi show that there is a correlation between gamma-ray luminosity and star formation rate (or infrared luminosity; \citealt{Fermi2012,Griffen2016,Linden2017}) and a correlation between gamma-ray luminosity and gas surface density \citep{Lacki2011}.  Both correlations imply that high star formation rate and high gas surface density systems approach the `proton calorimeter' limit \citep{Pohl1994,Thompson2007} in which most of the  CR proton energy is lost to pion decay before the CRs escape the galaxy.   Because the correlation between gamma-ray luminosity and infrared luminosity is better constrained than the correlation between gamma-ray luminosity and gas surface density, we use the former to calibrate the diffusion coefficients in our models:   $L_\gamma \simeq 2.3 \times 10^4 L_\odot (L_{TIR}/10^{9} L_\odot)^{1.25}$ \citep{Griffen2016} where $L_{TIR}$ is the total infrared luminosity from $0.1-1000 \, \mu m$.   

At high infrared luminosities, the infrared luminosity is linearly proportional to the star formation rate but this is not true for lower infrared luminosities where the ultraviolet radiation makes an increasingly important contribution to the total radiated starlight from galaxies.  We correct for this using \citet{Bell2003} who finds 
\be
\dot M_* \simeq 0.12 \mspy \left(\frac{L_{TIR}}{10^9 L_\odot}\right)\left(1 + \sqrt{\frac{10^9 L_\odot}{L_{TIR}}}\right)
\label{eq:mdotL}
\ee   
Given $L_\gamma(L_{TIR})$ and $L_{TIR}(\dot M_*)$, equation \ref{eq:Lgam} allows us to infer $t_{\rm diff}/t_{\pi}$ as a function of star formation rate.   This is shown with the purple line in the left panel Figure \ref{fig:gamma}:   $t_{\rm diff} \simeq t_{\pi}$ at $\dot M_* \sim 10^3 \mspy$ while $t_{\rm diff} \ll t_{\pi}$ for much lower star formation rates.   The latter is a direct consequence of the fact that $L_\gamma \ll A_\gamma L_*$ in systems like the Milky Way, M31, and the Magellanic clouds (e.g., \citealt{Fermi2012}) so that CR protons escape before losing most of their energy to pion decay.

\begin{figure}
\centering
\includegraphics[width=84mm]{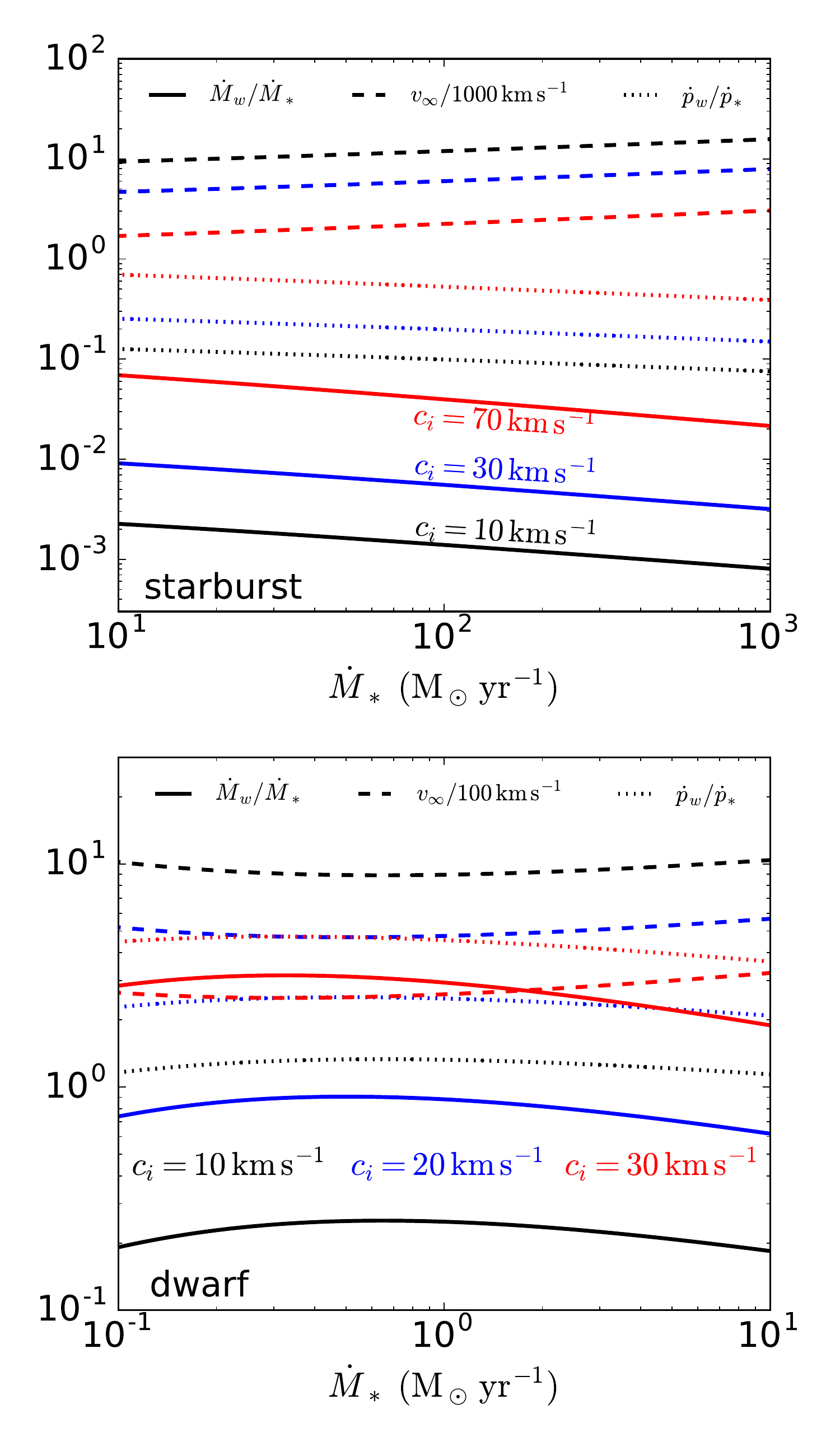}
\vspace{-0.4cm}
\caption{Empirically constrained mass-loss rates, terminal velocities, and momentum fluxes in galactic winds, as a function of galaxy star formation rate, for different values of the gas isothermal sound speed $c_i$.   The nuclear starburst model (top)  assumes $r_0=$ 0.3 kpc, $V_g=150 \kms$, and $\phi=1$ while the dwarf galaxy model assumes $r_0=1$ kpc, $V_g= 50\kms$, and $\phi=5$.  Note the different x-axis range and normalization of $v_\infty$ for the two panels.   The results for the star-forming disc galaxy model shown in Figure \ref{fig:gamma} are nearly independent of $c_i$ (see text) and so are not plotted here.   Starburst mass-loss rates are $\ll \dot M_*$ independent of $c_i$ while dwarf galaxy mass-loss rates can reach $\sim \dot M_*$ for larger values of $c_i$.  For starbursts with high star formation rates, the terminal speed and momentum flux are best interpreted as upper limits (see \S \ref{sec:pion} for details).}
\label{fig:gammaci}
\end{figure}

Given $t_{\rm diff}/t_{\pi}$ as a function of galaxy star formation rate (purple line Fig. \ref{fig:gamma}), we now use equation \ref{eq:tpi} to constrain the CR diffusion coefficient.   To do so, however, we need an estimate of the CR scale-height and gas surface density as a function of star formation rate.   We model $h_c = H_c/r_0$ using equation \ref{eq:hc} which assumes that the CR scale-height is set by diffusion in a CR-driven galactic wind.   We model the gas surface density $\Sigma_g$ using equation \ref{eq:SFR} which yields $\dot M_* \simeq \pi r_0^2 \sqrt{8} \pi G \Sigma_g^2 \phi /v_*$ where $r_0$ is the size of the galactic disc.   The free parameters of our model are thus the size of the star-forming galactic disc $r_0$, the galaxy circular velocity set by $V_g$, the gas isothermal sound speed $c_i$, and the dimensionless stellar contribution to the gravitational potential $\phi$ (as well as several `microphysics' parameters such as the CR energy per supernovae).  Given choices for these parameters, as well as the observed $L_\gamma-L_*$ correlation, we solve equation  \ref{eq:hc}, \ref{eq:SFR}, and \ref{eq:tpi} for $\kappa$, $h_c$, and $\Sigma_g$.   In what follows, we initially assume $c_i=10 \kms$ and consider parameters approximating 3 different classes of galaxies. Results for these classes are given in Figure \ref{fig:gamma}:  \begin{enumerate}
    \item Nuclear starbursts: black lines in Fig.~\ref{fig:gamma}; $r_0=$ 0.3 kpc, $V_g=150 \kms$, $\phi=1$. This model is meant to approximate the nuclear starbursts M82 and NGC 253  with $\dot{M}_*\simeq10$\,M$_\odot$ yr$^{-1}$, as well as ultra-luminous starbursts like Arp 220 with $\dot{M_*} \sim 10^2$\,M$_\odot$ yr$^{-1}$. 
    \item Dwarf galaxies: red lines in Fig.~\ref{fig:gamma}; $r_0=1$ kpc, $V_g= 50\kms$, $\phi=5$. This model is meant to approximate normal star-forming dwarf galaxies like the LMC or SMC and their potentially much more rapidly star-forming counterparts. 
    \item Star-forming disc galaxies: blue lines in Fig.~\ref{fig:gamma}; $r_0=3$ kpc, $V_g=150 \kms$, $\phi=5$. This model is meant to approximate the Milky Way, M31, and other  star-forming spirals in the local ($\dot M_* \sim \mspy$) and high-redshift ($\dot M_* \sim 10-100 \mspy$) universe.
\end{enumerate}
We discuss below (Fig. \ref{fig:gammaci} and eqs. \ref{eq:hcfit}-\ref{eq:pdotapprox2}) the scaling of our results to other gas sound speeds $c_i$.

The dotted lines in the left panel of Figure \ref{fig:gamma} show the  CR diffusion coefficient in units of $10^{29}$ cm$^2$ s$^{-1}$ required to explain the gamma-ray luminosities of star-forming galaxies, per the method described in the previous paragraph.   Figure \ref{fig:gamma} also shows the dimensionless CR scale-height $h_c$ we derive (dashed lines).   For the dwarf and starburst models in Figure \ref{fig:gamma}, $h_c \simeq 1$ and the diffusion coefficient is $\sim 2 \times 10^{29}$ cm$^2$ s$^{-1}$ $\gg r_0 c_i$ with only a modest factor of few variation with star formation rate.   For the star forming disc model, however, $h_c \sim 0.1$ and $\kappa \sim r_0 c_i \sim 10^{28}$ cm$^2$ s$^{-1}$. 
The dimensionless scale-height for our `disc' model is comparable to that inferred from synchrotron emission in nearby star-forming disc galaxies (e.g., \citealt{Krause2018}), though the latter depends on both the magnetic field and CR scale-heights.    
Our inferred values of $h_c \sim 0.1$ and $\kappa \sim 10^{28}$ cm$^2$ s$^{-1}$ for disc galaxies in Figure \ref{fig:gamma} are, however, factors of $\sim 5$ smaller than preferred in phenomenological models constrained by Milky Way CR data (e.g., \citealt{Trotta2011}). However, the estimated gamma-ray luminosity of the Milky-Way $\simeq 8 \times 10^{38}$ erg s$^{-1}$ \citep{Strong2010} is a factor of $\simeq 3$ lower than what would be predicted by its infrared luminosity given the correlations from \citet{Fermi2012} and \citet{Griffen2016} used here.  $h_c$ and $\kappa$ would then be 3 and 10 times larger, respectively (see eqs \ref{eq:hcfit} \& \ref{eq:kapapprox2} below), in better agreement with detailed Milky-Way modeling.  In addition, we show below that for $h_c \lesssim 1$, $h_c \propto c_i$ and $\kappa \propto c_i$ (eqs. \ref{eq:hcfit} \& \ref{eq:kapapprox2}).  Our disc galaxy model in Figure \ref{fig:gamma} would thus have a larger scale-height and diffusion coefficient if we assumed that the CRs primarily coupled to the warm-hot phase of the ISM, as is quite plausible. 

The key dimensionless number from \S \ref{sec:numerics} that determines the properties of CR driven galactic winds is $\kappa/(r_0 c_i)$;  winds accelerate rapidly and reach speeds significantly larger than $V_g$ if $\kappa/(r_0 c_i) \gtrsim 1$ (Fig. \ref{fig:flow_diff}).  For our 3 galaxy models in Figure \ref{fig:gamma}, we note that $r_0 c_i \simeq 10^{27}$ cm$^2$ s$^{-1}$ (starburst; black lines), $r_0 c_i \simeq 3 \times 10^{27}$ cm$^2$ s$^{-1}$ (dwarf; red lines), and $r_0 c_i \simeq 10^{28}$ cm$^2$ s$^{-1}$ (disc; blue lines).   A key conclusion from Figure \ref{fig:gamma} is that the gamma-ray data on star-forming galaxies are consistent with diffusion coefficients that are in the regime of $\kappa \gtrsim r_0 c_i$, though only marginally so for our star-forming disc model.  This does not prove that such diffusion coefficients are correct, at a minimum because it is possible that CR escape is not set by diffusion but rather by streaming or advection in winds driven by other mechanisms (e.g., supernovae), in which case our constraint on the CR diffusion coefficient using equation \ref{eq:Lgam} would not apply. But the diffusion coefficients inferred in Figure \ref{fig:gamma} are nonetheless a useful and instructive observational check on diffusive CR-driven galactic wind models.

Given the diffusion coefficients inferred in the left panel of Figure \ref{fig:gamma}, we can estimate the CR energy density in the galactic disc - which sets the base conditions for the wind - as follows:  if the star formation rate per unit area is $\dot \Sigma_*$, the CR pressure in the galaxy is given by (e.g., \citealt{Thompson2013b})\footnote{Note that this expression for $p_{c,0}$ is larger than its spherical counterpart used in \S \ref{section:analytic} (eq. \ref{eq:edotcdef}) by a factor of 2.   The difference is the surface area of a sphere ($4 \pi r_0^2$) vs. that of the bottom and top of the disk ($2 \times \pi r_0^2$).   We use the disk version in this section but our results are not sensitive to factor of 2 changes in $p_{c,0}$.}
\be
p_{c,0} \simeq \frac{\dot \Sigma_* E_{cr}}{6 H_c m_*} {\rm min}(t_{\rm diff},t_\pi)
\label{eq:pcequil}
\ee
where ${\rm min}(t_{\rm diff},t_\pi)$ determines the effective loss/escape time for CRs in the galaxy.  Using equation \ref{eq:SFR} and that fact that $t_{\rm diff} \lesssim t_\pi$ even at the highest star formation rates in Figure \ref{fig:gamma}, we find:
\be
\frac{p_{c,0}}{\pi G \Sigma^2_g \phi} \simeq \frac{\sqrt{2} E_{cr} h_c r_0}{3 \kappa p_*}
\label{eq:pcbase}
\ee 
The dashed lines in Figure \ref{fig:gamma} (middle panel) shows the resulting base CR pressures for our 3 galaxy classes/models.   For Milky Way-like ``star-forming disc" conditions with $r_0 \sim 3$ kpc, we find that $p_{c,0} \sim 0.1 \pi G \Sigma_g^2$, a bit smaller than local measurements in the Milky Way \citep{Boulares1990}, though we stress that our model is not intended to reproduce solar-circle measurements, but rather approximate the disc-averaged physical conditions. Figure \ref{fig:gamma} predicts that Milky Way-like galaxies have roughly the largest fraction of  disc pressure support from CRs, though for a given value of $r_0$, the base CR pressure only varies by a factor of a few over a factor of $\sim 10^4$ in star formation rate. The base CR pressure is, however, sensitive to $r_0$, and is significantly smaller, $\sim$ few $10^{-3}$, for nuclear starbursts like M82 and Arp 220.  Physically this is because for a given star formation rate, a smaller size for the star-forming disc implies higher gas densities and thus more rapid pion losses.   In order to be compatible with the gamma-ray observations, the diffusion time must be correspondingly shorter as well and thus the CR pressure cannot build to as large a value.   These conclusions are qualitatively similar to those of \citet{Lacki2010,Lacki2011} (see also \citealt{Crocker2021a,Crocker2020b}) who developed one-zone galaxy models with CRs in order to reproduce the far-infrared radio correlation.   They concluded that $p_{c,0} \sim \pi G \Sigma_g^2$ at low gas-surface densities, but that $p_{c,0} \ll \pi G \Sigma_g^2$ at the high gas densities of nuclear starbursts (see Fig. 15 of \citealt{Lacki2010} and section 6.3 of \citealt{Lacki2011}).

Finally, if we combine equation \ref{eq:pcbase} and equation \ref{eq:mdot3} we arrive at a simple expression for the mass-loss rate in CR-driven galactic winds:
\be
 \frac{\dot M_w}{\dot M_*} \simeq \frac{E_{cr}/m_*}{\Veffsq} \frac{c_i r_0}{3 \kappa} \left(\frac{4 h_c V_g^4}{c_{c,0}^2 c_i^2}\right)^{c_i^2/(2V_g^2)}
\label{eq:Mdotfin}
\ee
where $\sqrt{E_{cr}/m_*}$ is a velocity scale associated with CR feedback, which is $\simeq 220 \kms$ for $E_{cr} = 10^{50}$ erg and $m_* = 100 M_\odot$.

The solid lines in the middle panel of Figure \ref{fig:gamma} shows the mass-loading of CR-driven galactic winds $\dot M_w/\dot M_*$ from equation \ref{eq:Mdotfin} using $\kappa$ in the left panel of Figure \ref{fig:gamma} calibrated to gamma-ray observations.   We again show results for our 3 fiducial galaxy models.    The mass-loss rates in CR-driven winds are significant for a wide range of normal disc galaxy conditions with $\dot M_w \sim \dot M_*$, but are strongly suppressed ($\dot M_w \sim 10^{-3} \dot M_*$) in nuclear starbursts like M82, NGC 253, and Arp 220 because rapid CR diffusion and pion losses suppress the base CR pressure (Fig. \ref{fig:gamma}) and thus the mass-loss rate in the wind.   Dwarf galaxies lie somewhere in between with $M_w \sim 0.2 \dot M_*$.  

The right panel of Figure \ref{fig:gamma} shows the terminal velocity (eq. \ref{eq:vinfdiff}) and momentum flux $\dot p_w=\dot M_w v_\infty$ (in units of the photon momentum flux $\dot p_*=L/c$) for the same three galaxy models (compare with \citealt{Lochhaas2020}).  For the disc and dwarf models $v_\infty \sim 1000 \kms$ and $\dot p_w \sim \dot p_*$ while for the starburst model $v_\infty \sim 10^4 \kms$ and $\dot p_w \sim 0.1 \dot p_*$.   The large velocities and correspondingly lower momentum fluxes for the starburst model are a consequence of $\kappa \gg r_0 c_i$ needed to avoid overproducing the gamma-ray luminosities.

One of the uncertainties in assessing the implications of our results for observations is the appropriate value of the gas isothermal sound speed.   Most of the mass in the ISM is in cooler phases but most of the volume is in warmer phases.  As a result, it is plausible, though not guaranteed, that the warmer phases set the scattering rate and diffusion coefficient for the cosmic-rays.  Figure \ref{fig:gammaci} shows how the mass-loss rates, terminal velocities, and momentum fluxes we infer from gamma-ray data in our starburst and dwarf models depend on the assumed value of $c_i$.   We do not show similar results for the normal star-forming disc model because that model is in the regime $h_c \lesssim 1$ where the mass-loss rate, terminal velocity, and momentum flux given gamma-ray inferred diffusion coefficients are a very weak function of $c_i$ (e.g., the results in Figure \ref{fig:gamma} for the `disc' model apply to better than a factor of 2 accuracy for $c_i \lesssim 100 \kms$); this is derived analytically below.   Figure \ref{fig:gammaci} shows that the mass-loss rate increases notably with increasing $c_i$ for both our starburst and dwarf galaxy models.   The momentum flux increases more slowly with $c_i$ and the terminal velocity of the wind decreases due to the larger mass-loadings.   However, the qualitative conclusions drawn from Figure \ref{fig:gamma} remain robust.  Namely, for the starburst models $\dot M_w \ll M_*$ and for the dwarf models $\dot M_w$ is at most $\sim {\rm few \times} \, \dot M_*$.   The latter is, however, still  smaller than the large mass-loadings in dwarf galaxies typically needed to reconcile the galaxy stellar mass and dark matter halo mass functions (e.g., \citealt{Somerville2015}).

Luminous starbursts (e.g., Arp 220, and to a lesser extent M82) have gamma-ray luminosities approaching the calorimeter limit  $L_\gamma \simeq A_\gamma L_*$ due to $t_\pi \lesssim t_{diff}$ (e.g., \citealt{Lacki2011,Fermi2012,Griffen2016}).  In the calorimeter limit, the constraints on $\kappa$ in Figure \ref{fig:gamma} are best interpreted as upper limits, since the gamma-ray emission is roughly independent of $\kappa$ for $t_\pi \lesssim t_{diff}$.   In this regime, the estimated mass-loss rate is independent of $\kappa$ because the base CR pressure is set by $t_{\pi}$ in equation \ref{eq:pcbase} rather than $t_{diff}$.   However, because the asymptotic wind speed is $\propto \kappa^{1/2}$ (eq. \ref{eq:vinfapprox}), in the calorimeter limit, we can only empirically derive an upper limit on $v_\infty$ and the asymptotic wind kinetic energy and momentum flux.  Thus, particularly for the more luminous starbursts in Figures \ref{fig:gamma} \& \ref{fig:gammaci}, our results may be better interpreted as upper limits on the wind terminal velocity and momentum flux.  This further strengthens our conclusion that  winds due to cosmic-rays alone are weak in starburst galaxies and cannot drive their exceptional outflows (e.g., see \citealt{Barcos-Munoz2018})

We now derive analytic approximations to the results in Figures \ref{fig:gamma} and \ref{fig:gammaci}. These are valuable because they show how the results depend on all of the physical parameters of the problem.   In the analytics we assume that $r_s \sim r_0$ (see eq. \ref{eq:rc}), i.e., that the factor $\left(4 h_c V_g^4/[c_{c,0}^2 c_i^2]\right)^{c_i^2/(2V_g^2)} \sim 1$.  This is an excellent approximation for Figure \ref{fig:gamma} in which $c_i=10 \kms$, but is less applicable for the largest values of $c_i$ in Figure \ref{fig:gammaci}. In our analytic estimates, we also use a fit to our observational calibration of the CR diffusion timescale.  An approximate fit to the results in Figure \ref{fig:gamma} is given by
\be
\frac{t_{\rm diff}}{t_\pi} \simeq \alpha \ E_{cr,50}^{-1} \ \left(\frac{\dot M_*}{1 \, \mspy}\right)^{0.5} 
\label{eq:tpiobs}
\ee
with $\alpha \equiv 0.07 \alpha_{0.07} \simeq 0.07$.   Equation \ref{eq:tpiobs} is accurate to better than a factor of 2 over the entire range of $\dot M_*$ shown in Figure \ref{fig:gamma}.   It is not asymptotically correct, however, at either high or low star formation rates.   For high star formation rates, $L_\gamma \propto L_{TIR}^{1.25} \propto \dot M_*^{1.25}$.   Thus $L_\gamma/L_* \propto \dot M_*^{0.25}$ instead of $\propto \dot M_*^{0.5}$.  For low star formation rates, however, $L_{TIR} \propto \dot M_*^2$ so that $L_\gamma/L_* \propto \dot M_*^{1.5}$.  In practice,  we find that equation \ref{eq:tpiobs} is a good compromise, particularly given its simplicity.   In particular, following through the derivations of $h_c$, $\kappa$, $p_{c,0}$, and $\dot M_w/\dot M_*$ per equations \ref{eq:hc}, \ref{eq:tpi}, \ref{eq:pcbase}, and \ref{eq:Mdotfin} we can show how the results in Figure \ref{fig:gamma} depend on the various micro (CR and star formation) and macro (global galaxy) parameters in the problem.   Combining equations \ref{eq:hc}, \ref{eq:tpi}, \& \ref{eq:tpiobs} we find
\be
\begin{split}
h_c \simeq &  \min\bigg[1, \frac{0.9 \, E_{cr,50} v_{*,8.5}^{1/2}}{\alpha_{0.07} \, \phi^{1/2}} \left(\frac{c_i}{10 \kms}\right) \\ & \times \, \left(\frac{\rm kpc}{r_0}\right) \left(\frac{100 \kms}{V_g}\right)^2
  \bigg]
\label{eq:hcfit}
\end{split}
\ee
where $v_{*,8.5}=v_*/3000 \kms$.

There are two regimes depending on whether $h_c < 1$ or $h_c=1$ in equation \ref{eq:hcfit}.   For $h_c \sim 1$, which is the regime appropriate for dwarf galaxies and nuclear starbursts in our models in Figs \ref{fig:gamma} \& \ref{fig:gammaci},
\be
\kappa \simeq 3 \times 10^{29} \, {\rm cm^2 \, s^{-1}} \,  \left(\frac{h_c E_{cr,50} v_{*,8.5}^{1/2}}{\alpha_{0.07} \, \phi^{1/2}}\right) ,
\label{eq:kapapprox}
\ee
\be
\frac{p_{c,0}}{\pi G \Sigma_g^2 \phi} \simeq 0.03 \, \alpha_{0.07} \, \phi^{1/2}  v_{*,8.5}^{-3/2} \, m_{*,2}^{-1} \, \left( \frac{r_0}{3 \kpc}\right)
\label{eq:pbaseapprox}
\ee
\be
\begin{split}
\frac{\dot M_w}{\dot M_*} &  \simeq 0.1 \, \alpha_{0.07} \, \phi^{1/2} \, v_{*,8.5}^{-1/2} \, m_{*,2}^{-1} \\ & \times \left(\frac{100 \kms}{\Veff}\right)^2  \, \left(\frac{c_i r_0}{h_c 30 \kms \kpc}\right),
\label{eq:Mdotapprox}
\end{split}
\ee
where $m_{*,2}=m_*/100 M_\odot$.  Equations \ref{eq:kapapprox} and \ref{eq:vinfdiff} can also be combined to estimate the terminal velocity of the wind:
\be
\begin{split}
v_\infty & \simeq 1600 \, \kms \, \alpha_{0.07}^{-1/2} \, E_{cr,50}^{1/2} \, \phi^{-1/4} \, v_{*,8.5}^{1/4} \\ & \times \, \left(\frac{\Veff}{100 \, {\rm km s^{-1}}}\right) 
 \left(\frac{\rm 30 \, h_c \, kpc \, {\rm km \, s^{-1}}}{r_0 c_i}\right)^{1/2}
\end{split}
\label{eq:vinfapprox}
\ee
Equations \ref{eq:Mdotapprox} and \ref{eq:vinfapprox} then imply
\be
\begin{split}
\frac{\dot p_w}{\dot p_{*}} & \simeq 1 \, \frac{\alpha_{0.07}^{1/2} \, E_{cr,50}^{1/2} \, \phi^{1/4}}{m_{*,2} \, \epsilon_{*,-3.3}\, v_{*,8.5}^{1/4}} \\ & \times \, \left(\frac{100 \, {\rm km \,s^{-1}}}{\Veff}\right) 
 \left(\frac{r_0 c_i}{\rm 30 \, h_c \, kpc \, {\rm km \, s^{-1}}}\right)^{1/2}
 \end{split}
 \label{eq:pdotapprox}
 \ee

In the opposite regime of $h_c \lesssim 1$, which is the regime of star-forming disc galaxies in Figure \ref{fig:gamma}, we find
\be
\begin{split}
\kappa & \simeq 10^{29} \, {\rm cm^2 \, s^{-1}} \left(\frac{ E_{cr,50}^2 v_{*,8.5}}{\alpha_{0.07}^2 \, \phi} \right) \, \left(\frac{c_i}{10 \kms}\right) \\ & \times \, \left(\frac{100 \kms}{V_g}\right)^2 \, \left(\frac{3 \, \kpc}{r_0}\right),
\end{split}
\label{eq:kapapprox2}
\ee
\be
\frac{p_{c,0}}{\pi G \Sigma_g^2 \phi} \simeq 0.03 \, \alpha_{0.07} \, \phi^{1/2}  v_{*,8.5}^{-3/2} \, m_{*,2}^{-1} \, \left( \frac{r_0}{3 \kpc}\right)
\label{eq:pbaseapprox2}
\ee
\be
\frac{\dot M_w}{\dot M_*}  \simeq 0.2 \, \alpha^2_{0.07} \, \phi \, v_{*,8.5} \, m_{*,2}^{-1} \, \left(\frac{r_0}{3 \kpc}\right)^2,
\label{eq:Mdotapprox2}
\ee
\be
v_\infty \simeq 1000 \, \kms \, \alpha_{0.07}^{-1} \, E_{cr,50} \, \phi^{-1/2} \, v_{*,8.5}^{1/2} \, \left(\frac{3 \kpc}{r_0}\right)
\label{eq:vinfapprox2}
\ee
and 
\be
\frac{\dot p_w}{\dot p_{*}}  \simeq 1.3 \frac{\alpha_{0.07} \, E_{cr,50} \, \phi^{1/2} \, v_{*,8.5}^{3/2}}{m_{*,2} \, \epsilon_{*,-3.3}} \, \left(\frac{r_0}{3 \kpc}\right)
\label{eq:pdotapprox2}
\ee
We reiterate that $\phi$ in equations \ref{eq:hcfit}-\ref{eq:pdotapprox2} quantifies the contribution of stars and dark matter to the gravitational acceleration with $\phi \sim 1$ for gas-dominated systems and $\phi \sim 3-10$ for typical star-forming galaxies in the local Universe.    Also note that equations \ref{eq:pbaseapprox} and \ref{eq:pbaseapprox2} are identical, i.e. the same for $h_c < 1$ and $h_c=1$. This is because $p_c$ depends only on the ratio $H_c/\kappa$ (eq. \ref{eq:pcbase}) which is uniquely determined by our calibration of $t_{\pi}/t_{\rm diff}$ (eqs. \ref{eq:tpi} \& \ref{eq:tpiobs}).  

Equations \ref{eq:hcfit}-\ref{eq:pdotapprox2} do a good job of reproducing many key results shown in Figure \ref{fig:gamma}: (1) the characteristic diffusion coefficient $\sim 10^{28-29}$ cm$^2$ s$^{-1}$ relatively independent of galaxy properties required to reproduce the observed gamma-ray-star formation rate correlation, and (2) a base CR pressure $p_{c,0}$ and wind mass-loading $\dot M_w/\dot M_*$ that are relatively independent of star formation rate, but a strong function of the size of the star-forming disc $r_0$ at fixed star formation rate (because diffusion is much more rapid for smaller $r_0$ and decreases $p_{c,0}$ and $\dot M_w$).    
.   

Equations \ref{eq:hcfit}-\ref{eq:pdotapprox2} also elucidate how our gamma-ray inferences depend on the gas isothermal sound speed.   Figure \ref{fig:gamma} takes $c_i=10 \kms$.  Our star-forming disc model in Figure \ref{fig:gamma} is in the regime with $h_c \lesssim 1$, so that equations \ref{eq:kapapprox2}-\ref{eq:pdotapprox2} apply.   In this regime $\kappa \propto c_i$ so that $\kappa/r_0 c_i \sim 1$ from Figure \ref{fig:gamma} is in fact true relatively independent of $c_i$.   Moreover, for $h_c \lesssim 1$, $\dot M_w$ is independent of $c_i$ (eq. \ref{eq:Mdotapprox2}).  The results in the middle and right panels of Figure \ref{fig:gamma} are thus applicable over a wide range of $c_i$ (this is the reason that we do not plot the `disc' model in Fig. \ref{fig:gammaci}).  Our conclusion that CR-driven winds constrained by gamma-ray observations of  disc galaxies are dynamically important with $\dot M_w \sim \dot M_*$ is thus robust to uncertainties in $c_i$.   

In contrast to the disc model, the dwarf and starburst models in Figure \ref{fig:gamma} are in the regime where $h_c \sim 1$ so that equations \ref{eq:kapapprox}-\ref{eq:pdotapprox} apply.  In this case the inferred $\kappa$ is independent of $c_i$ (eq. \ref{eq:kapapprox}); however, even $c_i \sim 100 \kms$ still corresponds to $\kappa/r_0 c_i \gtrsim 1$ so that the qualitative physics of the wind does not change.  Equations \ref{eq:Mdotapprox}, \ref{eq:vinfapprox}, \& \ref{eq:pdotapprox} predict that
the inferred mass-loss rate, terminal velocity, and momentum flux given gamma-ray constraints scale $\propto c_i, c_i^{-1/2}$ and $c_i^{1/2}$, respectively.  This is consistent with, though weaker than, the trends in Figure \ref{fig:gammaci}. The stronger dependence on $c_i$ in Figure \ref{fig:gammaci} is because for larger values of $c_i$ the approximation $\left(4 h_c V_g^4/[c_{c,0}^2 c_i^2]\right)^{c_i^2/(2V_g^2)} \sim 1$ used to derive equations \ref{eq:kapapprox}-\ref{eq:pdotapprox2} is no longer as accurate.   Physically, this is because for larger values of $c_i$ and smaller $c_{c,0}$ the sonic point $r_s$ is at somewhat larger radii $\sim {\rm few \times} r_0$, which increases $\dot M_w$ (eq. \ref{eq:mdot3}) and decreases $v_\infty$ (eq. \ref{eq:vinfdiff}).

We conclude this section by reiterating that our most general expressions for the base CR pressure $p_{c,0}$, mass-loss rate $\dot M_w$, terminal velocity $v_\infty$, and momentum flux $\dot p_w$ in CR-driven galactic winds are given in \S \ref{section:analytic}. Those results depend, however, on a theoretically and observationally uncertain CR diffusion coefficient.  In this section (and in Figs \ref{fig:gamma} \& \ref{fig:gammaci}) we have estimated the CR diffusion coefficient $\kappa$ using the existing (but limited) gamma-ray data from Fermi on pion-decay in star-forming galaxies, thus enabling more concrete predictions of the properties of CR-driven galactic winds.   

\section{Summary and Discussion}
\label{sec:disc}

 The physics of cosmic ray (CR) transport in galaxies and in the circumgalactic medium remains a significant uncertainty in assessing the impact of CRs on galaxy formation.  A central question is what determines the scattering mean free path of CRs, and how this depends on local plasma conditions (e.g., \citealt{Amato2017,Hopkins2020b}). In this paper, we have assumed that CR transport can be modeled by a spatially independent diffusion coefficient.  The diffusion approximation for CR transport is particularly appropriate if ambient turbulence scatters the CRs (vs. scattering by fluctuations excited by the CRs themselves).   A companion paper will consider the case of CR transport mediated by the streaming instability.  These two mechanisms of CR transport differ dramatically in their predictions for how the CR pressure decreases away from a galaxy:  in the limit of rapid CR diffusion, $p_c \propto r^{-1}$ (eq. \ref{eq:pcsol}), i.e., the CR pressure scale-height is of order the size of the system, while in the limit of rapid CR streaming, $p_c \propto \rho^{2/3}$ (for a split-monopole field geometry; e.g., \citealt{Mao2018}) and so the CR pressure scale-height is tied to that of the gas.  This difference in the dynamics of the CRs in general leads to significantly different wind properties for the two CR transport models, as has been highlighted previously in numerical simulations (e.g., \citealt{Wiener2017,Chan2019}).   One aim of this paper and its companion is to understand these differences analytically and using idealized time-dependent numerical simulations, thus elucidating how the properties of CR-driven galactic winds depend on global galaxy properties and the physics of CR transport.

 In this paper, we analytically estimated the properties of galactic-winds driven by diffusion by assuming that the CR diffusion timescale is short compared to the flow time (or dynamical time) near the base of the wind; this requires CR diffusion coefficients $\kappa \gtrsim r_0 c_i$ where $r_0$ is the size of the galaxy (i.e., the star-forming disc) and $c_i$ is the gas sound speed.   In this limit, the asymptotic kinetic energy flux carried by the wind is comparable to that supplied to the CRs at the base of the wind, i.e., the wind is energy conserving.  The mass-loss rate of CR driven winds has the form $\dot{M}_w \sim 2 \pi r_0^2\,\rho_0\, c_i \, (c_{c,0}/V_g)^2 \sim 2 \pi r_0^2 p_{c,0} c_i/V_g^2$ (eq. \ref{eq:mdot3}; see Fig.~\ref{fig:mdotdiff}), and the asymptotic wind speed is $V_\infty\simeq 2 V_g (3 \kappa/r_0 c_i)^{1/2}$ (eq. \ref{eq:vinfdiff})  where $\rho_0$, $p_{c,0}$ and $c_{c,0}$  are the gas density, CR pressure, and CR sound speed at the base of the outflow and  $\sqrt{2}V_g$ is the rotation velocity of the galaxy. Equation \ref{eq:mdotf} compares this estimate of the mass-loss rate in CR-driven winds to the galaxy star formation rate, with $\dot M_w/\dot M_* \propto 1/\kappa$.  Physically, for a given rate of CR production, set by the star formation rate, the CR pressure in the galaxy, and thus the strength of the wind, decreases with increasing diffusion coefficient since the CRs escape the galaxy more rapidly.

 In addition to our analytic estimates, we also carried out time-dependent spherically symmetric simulations of CR-driven winds using the two-moment CR transport scheme for {\tt Athena++} developed by \citet{Jiang2018}.   The simulations show that, for $\kappa \gtrsim r_0 c_i$, the analytic estimates for the mass-loss rate, terminal speed, and CR scale-height near the base of the wind are accurate to $\sim 50\%$ over a factor of $\sim 30$ in CR diffusion coefficient, $\sim 30$ in base CR pressure, and $\sim 100$ in the ratio of the escape speed to the gas sound speed (Fig. \ref{fig:mdotfin}; see Table \ref{tab:compare} for the full range of simulations).   In addition, the simulations show that there is a critical value of the CR diffusion coefficient $\kappa \simeq r_0 c_i$ below which the character of the solution changes considerably.   For $\kappa \lesssim r_0 c_i$, CR-driven winds accelerate much more slowly and are nearly hydrostatic over a very extended radial range.  In this regime most of the energy supplied to CRs at the base of the wind goes into work against gravity expanding to large radii (Fig. \ref{fig:Edot_diff_lowkap}): the asymptotic kinetic energy flux in the wind is only a small fraction of that initially supplied to the CRs (see the last column of Table \ref{tab:compare}).  These low $\kappa$ solutions are CR analogues of photon-tired stellar winds \citep{Owocki1997}.   The mass-loss rate in this regime can be accurately estimated from global energy conservation as $\dot M_w \simeq \dot M_{max} \simeq 2 \dot E_c/v_{esc}^2$ (eqs. \ref{eq:mdotmax} \& \ref{eq:mdotmaxsim}), where $\dot E_c$ is the energy per unit time supplied to CRs at the base of the wind.   This maximum possible mass-loss rate in CR-driven winds is quite large, $\simeq \dot M_* (300 \, \kms /v_{esc})^2$ (eq. \ref{eq:mdotmaxvsmdotstar}).  For $\kappa > r_0 c_i$, however, the actual outflow rate is much less than this maximal value (eq. \ref{eq:mdotwvsmdotmax}).
 
A key difference between our treatment of CR-driven winds in this paper and analogous treatments of stellar winds driven by radiation in the diffusion approximation (e.g., \citealt{Owocki2017}) is that  stellar wind theory is typically formulated in terms of a given photon-matter cross section $\sigma$, which sets the Eddington luminosity.  By contrast, here we are considering a fixed CR diffusion coefficient, equivalent to a fixed value of the mean-free path $1/(\sigma \rho)$.  This difference means that many solutions in stellar wind theory do not directly carry over to the CR problem, although many of the important concepts do.  

In our models with $\kappa \gtrsim r_0 c_i$, the properties of CR-driven winds are largely set close to the `base' of the wind, i.e., near the galaxy.   In particular, the sonic point - which sets the mass-loss rate - is close to the base of the wind (eq. \ref{eq:rc}) unless $c_i \sim V_g$ and the energy flux in the wind - which sets the terminal velocity - is set by the CR diffusive flux at the base (eq. \ref{eq:vinfdiff} and associated discussion).   As a result, we suspect that the properties of these solutions are unlikely to be that sensitive to spatial variation in the CR diffusion coefficient unless there are large variations at small radii near the sonic point.   By contrast, our solutions with $\kappa \lesssim r_0 c_i$ accelerate much more slowly (Fig. \ref{fig:Edot_diff_lowkap}) and are likely much more sensitive to spatial variation in the microphysics of CR transport.  In addition, because the low $\kappa$ solutions have a kinetic power $\dot E_k$ at large radii that is small compared to the cosmic ray power at the base of the wind, they are likely more sensitive to the ambient pressure in the CGM, which could confine lower $\dot E_k$ outflows.

 Our time-dependent simulations allow us to study the stability of the analytic steady state wind solutions.   Nearly all of our simulations reach a laminar steady state with no evidence of instability.   This is at first glance surprising since \citet{Drury1986} showed that CR diffusion in the presence of a background CR pressure gradient renders sound waves linearly unstable.   We show in Appendix \ref{sec:appendixA}, however, that the growth rate of the sound wave instability is not fast enough compared to the flow time in the wind for the instability to grow significantly; the one exception to this is our lowest gas sound speed simulation (the $V_g=200 c_i$ simulation in Table \ref{tab:compare}; see Figure \ref{fig:app}).    Appendix \ref{sec:appendixA} also carries out a WKB linear stability calculation (neglecting the background cosmic-ray pressure gradient) for the two-moment CR transport scheme used in our simulations, and shows that sound waves and entropy modes are linearly stable in the presence of CR diffusion, consistent with the steady state solutions found in the simulations.   
 
 A key parameter that sets the strength of the galactic wind in our models is the CR pressure in the bulk of the ISM (with $\dot M_w \propto p_{c,0}$).   If $p_{c,0} \sim \pi G \Sigma_g^2 \phi$ (the pressure required for hydrostatic equilibrium), CR-driven winds will have dynamically important mass-loss rates with $\dot M_w \gtrsim \dot M_*$ (eq. \ref{eq:mdot0}).   If, however, $p_{c,0} \ll \pi G \Sigma_g^2 \phi$, then since $\dot M_w \propto p_{c,0}$ (eq. \ref{eq:mdot3}), the mass-loss rates will be significantly smaller.   The equilibrium CR pressure $p_{c,0}$ is in turn set by CR escape (i.e., the diffusion coefficient $\kappa$) and/or hadronic losses (eq. \ref{eq:pcequil}). To assess the implications of our results for the role of CRs in driving galactic winds, it is thus necessary to estimate the CR diffusion coefficient in other galaxies.  This remains a daunting task from first principles, so we instead turned to observations (see \S \ref{sec:pion}).    In particular, observations of the non-thermal emission from CRs in other galaxies provide direct constraints on CR diffusion coefficients and the CR pressure in galaxies (e.g., \citealt{Lacki2010,Lacki2011,Crocker2021a}).   The non-thermal gamma-ray emission from neutral pion decay is particularly important in this regard because (1)  it constrains the properties of CR protons (vs. synchrotron emission), and (2)  observations at GeV energies by Fermi, though modest in number, directly constrain the CRs that dominate the total CR pressure.   In \S \ref{sec:pion} we developed a simple analytic model interpreting gamma-ray observations in the context of diffusive CR transport.  This model essentially derives the theoretically and observationally uncertain diffusion coefficient as a function of the observed gamma-ray luminosity of galaxies.   We find that a model with a diffusion coefficient $\sim 10^{28-29}$ cm$^2$ s$^{-1}$ (Fig. \ref{fig:gamma} and eqs. \ref{eq:kapapprox} \& \ref{eq:kapapprox2})  is consistent with the Fermi data on gamma-ray emission from star-forming galaxies.  This is consistent with similar estimates by \citet{Chan2019} and \citet{Hopkins2020a} and their more detailed numerical calculations.
 
 Our constraint on the diffusion coefficient in other galaxies also translates into an estimate of the CR pressure in galactic discs.  For typical star forming galaxies with disc sizes $r_0 \sim 3$ kpc, we find that the CR pressure is of order 10\% of the pressure required for vertical hydrostatic equilibrium in the disc (Fig. \ref{fig:gamma} and eq. \ref{eq:pbaseapprox}).  This is reasonably consistent with Milky Way measurements.   However, the fractional contribution of CRs to pressure support in the disc is $\propto r_0$ and is only $\sim 10^{-2.5}$ for typical nuclear starburst conditions (Fig. \ref{fig:gamma} and eq. \ref{eq:pbaseapprox}).   Physically, this is because in more compact star-forming regions, the gas densities are higher and thus pion losses are stronger.  In addition, the CR diffusion time is shorter.   There is thus less time for the CR pressure to build up and so the equilibrium CR pressure in the disc is lower.   These conclusions are consistent with the earlier work of \citet{Lacki2010} based on modeling the far infrared-radio correlation.   

 Our results on the CR diffusion coefficient and CR pressure implied by gamma-ray observations can be used to estimate the properties of CR-driven galactic winds across a wide range of galaxies.    The middle panel of Figure \ref{fig:gamma} plots the resulting ratio of the CR-driven mass-loss rate to the star formation rate for three fiducial galaxy models, while the right panel shows the terminal velocity and momentum flux of the wind.  We find that for massive star-forming disc galaxies, the mass-loss rates are of order the star formation rate, momentum fluxes are of order $\dot p_* = L/c$, and terminal velocities are $\sim 500 \kms$ (a few times the circular velocity).   For lower-mass dwarf galaxies, however, we find that CRs are somewhat less efficient at driving winds ($\dot M_w \sim 0.2 \dot M_*$ and $\dot p_w \sim L/c$), primarily because the CR diffusion time is so short (to explain the gamma-ray data) that the CR pressure in the disc is comparatively low.   This is even more true in
 nuclear starbursts:   CRs become much less efficient at driving winds with $\dot M_w/\dot M_* \propto r_0$ (eq. \ref{eq:Mdotapprox}) and $\dot M_w/\dot M_* \sim 10^{-2}-10^{-3}$ for well-studied local starbursts like M82 and Arp 220 (Fig. \ref{fig:gamma}).    This conclusion fundamentally rests on our inference that $p_{c,0} \ll \pi G \Sigma_g^2 \phi$ given CR diffusion coefficients and pion loss timescales motivated by gamma-ray observations.   An independent observational probe of the CR proton pressure in other galaxies would be a valuable test of our models.
 
One of the uncertain parameters in applying our results to observations is the appropriate isothermal gas sound speed.  This depends on the phase of the ISM that the cosmic-rays  most effectively couple to.  Figure \ref{fig:gamma} assumes $c_i = 10 \kms$, which is an appropriate mass-averaged value in the Milky Way.  For typical star-forming disc galaxy parameters, we find that the properties of the winds using gamma-ray constrained diffusion coefficients are weakly dependent on $c_i$ (also derived analytically in equations \ref{eq:Mdotapprox2}-\ref{eq:pdotapprox2}). Our conclusion that cosmic-rays are  a significant source of winds in normal disc galaxies is thus reasonably robust to the uncertainty of the phase of the ISM that primarily determines CR transport.  

Figure \ref{fig:gammaci} shows our gamma-ray inferred wind properties for dwarf galaxy and nuclear starburst models for larger values of $c_i$, appropriate if cosmic-rays primarily couple to volume filling warm-hot gas.   For these galaxy models, the mass-loss rate can increase significantly for larger values of $c_i$, as does the momentum flux in the wind; the terminal speed of the wind is correspondingly smaller for larger $c_i$.     However, our general conclusions are reasonably robust to uncertainties in $c_i$:   the mass-loss rates due to CRs alone in starburst galaxies are $\ll \dot M_*$ and in dwarf galaxies are at most $\sim {\rm few} \times \dot M_*$.  The latter is still below what is typically needed to reconcile the stellar and dark matter halo mass functions (see, e.g., \citealt{Muratov2015} Table 3).

As noted earlier in the discussion, the maximum mass-loss rate in CR-driven winds allowed by energy conservation is appreciable, $\dot M_{\rm max} \simeq \dot M_* (300 \, \kms /v_{esc})^2$ (eq. \ref{eq:mdotmaxvsmdotstar}).  Mass-loss rates $\sim \dot M_{\rm max}$ would be particularly important in dwarf galaxies.   However, these large mass-loss rates are only realized when the outflow is very slow and most of the energy supplied to CRs by star formation goes into work leaving the gravitational potential of the galaxy (Fig. \ref{fig:Edot_diff_lowkap}).  This is turn requires low CR diffusion coefficients.  Such slow outflows would produce gamma-ray luminosities in dwarf galaxies and compact nuclear starbursts larger than are observed.   This is the fundamental observational constraint that leads us to favor larger diffusion coefficients and modest mass-loss rates in dwarf and starburst galaxies.    A corollary of this result is that in all of our models calibrated to explain gamma-ray luminosities well below the proton-calorimeter value, most of the CR proton energy is vented into the CGM.   Even if the CR-driven mass-loadings on galactic scales are modest,  CRs may play an important `preventive' feedback role on CGM scales and/or may significantly modify the dynamics and thermodynamics of the CGM (as was indeed found in the simulations of \citealt{Ji2020}).

It is instructive to  compare our results to related results in the literature.   For example, \citet{Booth2013} assumed $\kappa = 3 \times 10^{27}$ cm$^2$ s$^{-1}$ 
in their simulations of the impact of cosmic-rays on star-forming galaxies.   By contrast, \citet{Salem2014} considered a range of diffusion coefficients $\kappa = 3 \times 10^{27}-10^{29}$ cm$^2$ s$^{-1}$ in a similar study.   Neither work compared to gamma-ray observations.   Our results strongly disfavor the low diffusion coefficient used by \citet{Booth2013} and favor the upper end of the values modeled in \citet{Salem2014}.    \citet{Chan2019} studied three-dimensional simulations of idealized galaxies with CRs and other forms of stellar feedback, and directly compared to gamma-ray observations.  They also concluded that CR diffusion coefficients of $\sim 10^{29}$ cm$^2$ s$^{-1}$ were required for consistency with gamma-ray observations.  \citet{Hopkins2020a} reached similar conclusions using cosmological zoom-in simulations.  Our analytics help firm up the conclusions drawn from these simulations and show how they depend on other  stellar feedback parameters and the galaxy model (see, in particular, our analytic scalings in equations \ref{eq:kapapprox}-\ref{eq:pdotapprox2}).    Both \citet{Chan2019} and \citet{Hopkins2020a} also found, as we do, that while CRs can drive winds in Milky-way mass galaxies, CRs are not very important wind-drivers in dwarf galaxies relative to other mechanisms.   

A significant difference between our solutions and the cosmological zoom-in simulations with CRs of  \citet{Hopkins2020a}, \citet{Ji2020}, and \citet{Hopkins2021} is that we find that advection of CR energy by the gas motion becomes the dominant CR transport mechanism relatively close to the base of the wind, with the gas kinetic energy flux taking over at yet larger radii (see Fig. \ref{fig:Edot_diff} and eq. \ref{eq:radv}).    By contrast,   \citet{Hopkins2020a}, \citet{Ji2020}, and \citet{Hopkins2021} argue that diffusion sets up a $p_c \propto r^{-1}$ profile throughout the CGM.  A possible resolution of this difference is that diffusion would likely again be the dominant CR transport mechanism exterior to a termination shock between a galactic wind and the CGM, which is not included in our calculations. It is also worth noting that our simulations require high resolution to resolve the acceleration of the gas at small radii, particularly for colder phases of the ISM, i.e., larger $V_g/c_i$ (see Table \ref{tab:compare}).  This  is not achievable in cosmological simulations.  If we take our fiducial $\kappa = 10$, $V_g=10$ simulation (Table \ref{tab:compare}) and reduce the resolution to $dr/r=0.05$, the mass-loss rate increases by a factor of $\sim 4$.   This is, however, almost certainly boundary condition dependent, and it is not clear how this result would change for cosmological simulations which do not have any boundary in the galaxy. 

Finally, we stress that our observational calibration of CR diffusion coefficients using Fermi gamma-ray data is based on a limited sample of galaxies, primarily those in the local group, M82, NGC 253, and Arp 220 \citep{Fermi2012,Griffen2016}.  It is thus entirely possible that there are physical correlations of CR transport with galaxy properties (gas density, metallicity, galaxy size, ...) that are not revealed by the current data.   Despite this caveat, given the particularly large theoretical uncertainties in the microphysics of CR transport, we believe that observational calibration of the models is an important constraint, and one that will hopefully improve in the coming years.
 
\section*{Data Availability} The numerical simulation results used in this paper will be shared on reasonable request to the corresponding author.  
 \vspace{-0.3cm}
 \section*{Acknowledgments} We thank Andrea Antoni, Phil Hopkins, Philipp Kempski, S. Peng Oh, Eve Ostriker, and Jono Squire for useful conversations.  EQ thanks the Princeton Astrophysical Sciences department and the theoretical astrophysics group and Moore Distinguished Scholar program at Caltech for their hospitality and support.  EQ was supported in part by a Simons Investigator Award from the Simons Foundation and by NSF grant AST-1715070.   The Center for Computational Astrophysics at the Flatiron Institute is supported by the Simons Foundation. TAT is supported in part by NSF grant \#1516967 and NASA grant \#80NSSC18K0526. TAT acknowledges support from a Simons Foundation Fellowship and an IBM Einstein Fellowship from the Institute for Advanced Study, Princeton, while a portion of this work was completed.This research made extensive use of Matplotlib \citep{Hunter:2007} and Astropy,\footnote{http://www.astropy.org} a community-developed core Python package for Astronomy \citep{Astropy1, Astropy2}.

\bibliography{ref}

\vspace{-0.2cm}
\begin{appendix}
\section{Linear Stability}
\label{sec:appendixA}
In this Appendix, we study the linear stability of the CR magnetohydrodynamic equations.  Since the simulations are one-dimensional we restrict ourselves to one dimensional perturbations and also consider a local Cartesian approximation instead of the global spherical geometry used in \S \ref{sec:numerics}.    Physically, this system of equations admits longitudinal sound waves, in which both gas and CR pressure are the restoring force, as well as gas and CR entropy modes.   In what follows, we show that ignoring background gradients, the two-moment CR system is linearly stable in the presence of cosmic-ray diffusion.    There is, however, an instability driven by a background cosmic-ray pressure gradient that is present in the one-moment CR system \citep{Drury1986}, i.e., the instability does not rely on the finite speed of light.  We show, however, that this instability grows too slowly to be dynamically important in galactic winds, consistent with the laminar numerical solutions we found in \S \ref{sec:numerics}.  The only exception to this is if the gas sound speed is very low, as we show in Figure \ref{fig:app}.   
\vspace{-0.3cm}
\subsection{Instabilities of the Two-Moment System}
\label{sec:lin2mom}

We assume here that perturbations are $\propto \exp(-i \omega t + i k r)$ and that $k H \gg 1$ (where $H$ is a characteristic length-scale in the equilibrium state) so that a WKB analysis is appropriate.   For now, we neglect the background gradients in the problem.

The key frequencies in the problem are the isothermal gas sound wave frequency
\be
\omega_g = k c_i,
\label{eq:omg}
\ee
the adiabatic CR sound wave frequency
\be 
\omega_c = k c_{\rm eff} \equiv k \sqrt{4 p_c/3 \rho},
\label{eq:omc}
\ee
the CR diffusion frequency associated with the assumed constant diffusion coefficient $\kappa$ 
\be
\omega_d = k^2 \kappa
\label{eq:omd}
\ee
and a characteristic frequency in the problem due to the finite speed of light, which we define as
\be
\omega_M = \frac{v_M^2}{\kappa} 
\label{eq:omM}
\ee
Note that in the simulations described in \S \ref{sec:numerics}, $\kappa \sim 1-30 c_i r$ so that $\omega_d/\omega_g \sim k r (\kappa/c_i r) \gg 1$.   The same inequality holds for $\omega_d/\omega_c$.   By contrast, $\omega_d/\omega_M \sim \kappa^2 k^2/v_M^2 \sim (k \ell)^2 < 1$ is required for the fluid approximation to the CR dynamics to be valid (where $\ell$ here is the CR mean free path).   

Working in the WKB limit, the one-dimensional linear dispersion relation for equations \ref{eq:CR2mom} is given by
\be
\label{eq:DRdiff}
\begin{split}
0  = &  \frac{3 \omega^4}{\omega_M} + i \omega^3\left(1 + 3 \frac{\omega_c^2}{\omega_d \omega_M}\right) - \omega^2 \left(\omega_d + 3 \frac{\omega_g^2}{\omega_M}\right) \\
-  & \ i \omega (\omega_g^2 + \omega_c^2) +  \omega_g^2 \omega_d
\end{split}
\ee

In the rapid diffusion limit of $\omega_d \gg \omega_g, \omega_c$ and $\omega_M \gg \omega_d$, the 4 solutions to equation \ref{eq:DRdiff} are all stable:
\be
\label{eq:rapid_diff}
\begin{split}
& \omega  \simeq \, -i  \, \frac{\omega_M}{3} \\
& \omega  \simeq \, -i \, {\omega_d}\\
& \omega  \simeq \pm \omega_g -i \, \frac{\omega_c^2}{2 \omega_d}
\end{split}
\ee
The first two solutions in equation \ref{eq:rapid_diff} are strongly damped entropy modes.   The last is a weakly damped gas sound wave.  Physically, the latter wave arises because in the limit $\omega_d \rightarrow \infty$, CR pressure gradients are completely wiped out by  diffusion and the only restoring force for a sound wave is the gas pressure.   At finite $\omega_d$, there is a small residual CR pressure gradient, the diffusion of which leads to damping of the associated sound wave.   

We reiterate that the rapid diffusion ordering used to derive equation \ref{eq:rapid_diff} is the appropriate one for our simulations in \S \ref{sec:numerics}.  The absence of any growing modes in equation \ref{eq:rapid_diff} is consistent with the numerical solutions which find laminar wind solutions.   

In the limit of slow CR diffusion, $\omega_d \ll \omega_g, \omega_c \ll \omega_M$ the solutions of equation \ref{eq:DRdiff} are also damped, namely
\be
\label{eq:slow_diff}
\begin{split}
& \omega  \simeq \, -i  \, \frac{\omega_M}{3} \\
& \omega  \simeq \, -i \, {\omega_d} \frac{\omega_g^2}{\omega_g^2 + \omega_c^2} \\
& \omega  \simeq \pm \sqrt{\omega_g^2 + \omega_c^2} -i \, \frac{\omega_d}{2} \frac{\omega_c^2}{\omega_g^2 + \omega_c^2}
\end{split}
\ee
\vspace{-0.2cm}
\subsection{Instabilities of the One-Moment CR System with Background Gradients}
\label{sec:lin1mom}
Instabilities of the one-moment CR system for a homogeneous background can be derived using the results in \S \ref{sec:lin2mom} by taking $v_M \rightarrow \infty$. The sound and entropy modes are both stable in this limit.   Including background gradients in the calculation, however, leads to an instability of sound waves that was discussed by \citet{Drury1986}. We briefly summarize a derivation of this instability for completeness and then discuss its relevance to our galactic wind simulations.   The \citet{Drury1986} instability is present in the one-moment CR system and so we restrict our analysis to this limit for ease of algebra.  

We consider an isothermal gas plus CR system that satisfies the following conservation laws 
\be
\frac{\partial \rho}{\partial t} + \frac{d \rho v}{d z}  = 0
\label{eq:massApp}
\ee
\be
\rho \frac{\partial v}{\partial t} + \rho v \frac{d v}{d z} = - c_i^2 \frac{d \rho}{d z} - \frac{d p_c}{d z} - \rho g
\label{eq:momApp}
\ee
We linearize equations \ref{eq:massApp} \& \ref{eq:momApp}.   To start we assume that all perturbations, labeled by $\delta$, are $\propto \exp(-i \omega t)$ but we do not Fourier transform in z.   We do the latter only at the end of the calculation to ensure that all background gradient terms are properly kept.    The linearly perturbed equations are then
\be 
i \omega \delta \rho = \frac{d (\rho \delta v)}{d z}
\label{eq:linmass}
\ee
\be
-i \omega \rho \delta v = -c_i^2 \frac{d \delta \rho}{d z} - \frac{d \delta p_c}{d z} - \delta \rho g
\label{eq:linmom}
\ee
Equations \ref{eq:linmass} and \ref{eq:linmom} can be combined to yield
\be
\omega^2 \delta \rho = -c_i^2 \frac{d^2 \delta \rho}{d^2 z} - \frac{d^2 \delta p_c}{d^2 z} - g \frac{d \delta \rho}{d z} 
\label{eq:comb}
\ee

In the limit of rapid CR diffusion, the linearized CR energy equation with diffusion (eqs. \ref{cr_energy} and \ref{equilibrium_flux}) simply becomes $\kappa \, d^2 \delta p_c/d^2 z \simeq 0$.    Substituting this into eq \ref{eq:comb}, assuming perturbations $\propto \exp[i k z - z/(2H)]$, where $H$ is the density scale-height, and using hydrostatic equilibrium in the background yields 
\be
\omega = |k c_i| + i \frac{ c_{\rm c}^2}{2 c_i} \frac{k}{|k|} \frac{d \ln p_c}{dz}
\label{eq:df}
\ee
to $O(1/H)$ ($c_{\rm c}^2 = p_c/\rho$ as in the main text).   Equation \ref{eq:df} is equivalent to the dispersion relation in \citet{Drury1986} in the limit of rapid CR diffusion.   \citet{Drury1986} further show that the rapid diffusion approximation leading to equation \ref{eq:df} only applies if $\kappa \gtrsim 4/3 |d\ln p_c/dr|^{-1} c_i$; otherwise the system is stable.   Our lowest $\kappa$ simulations in Table \ref{tab:compare} with $\kappa/r_0 c_i = 0.33, 0.11$ are stable at most radii per this condition; otherwise, the rapid diffusion approximation is a good one in our simulations.   

The number of e-foldings for the \citet{Drury1986} instability can be estimated as $A(r) \simeq {\rm Im}(\omega) H_\rho/c_i$ where $H_\rho$ is the density scale-height on which the background structure changes and the flow accelerates.  Using equations \ref{eq:df} and \ref{eq:hc}, we find
\be
A(r) \sim \sqrt{\frac{r_0 c_i}{\kappa}}\frac{c_c^2}{c_i V_g}
\label{eq:amp}
\ee
near the base of the outflow where the instability derivation is applicable.   For our fiducial simulation with $\kappa \sim 10 r_0 c_i$ and $c_c \simeq c_i \simeq 0.1 V_g$ near the base, we find $A \sim 0.03$, i.e., very little growth of the instability.  This is consistent with our laminar numerical simulations.   Fundamentally, the reason for this is that the CR pressure gradient that drives the \citet{Drury1986} instability is very shallow in galactic winds driven by CR diffusion, with a CR pressure scale-height much larger than the density scale-height in the subsonic portion of the wind at small radii where equation \ref{eq:df} applies (see Fig. \ref{fig:flow_diff}). The large CR pressure scale-height in the present context means that the the growth of the \citet{Drury1986} instability is slow and is the key reason why nearly all of our simulations do not show any sign of this linear instability.   

\begin{figure}
\centering
\includegraphics[width=88mm]{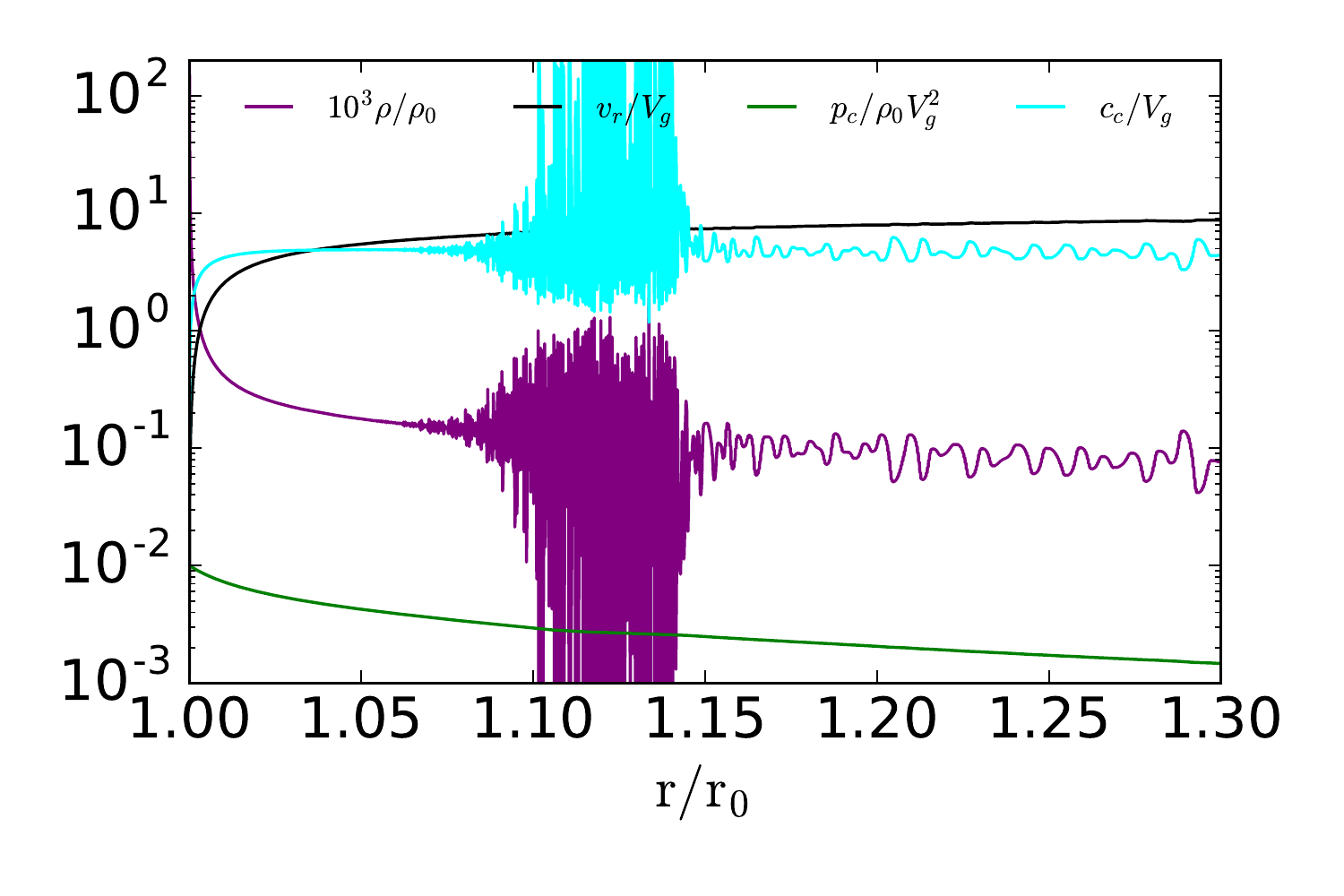}
\vspace{-0.4cm}
\caption{Density, velocity,  cosmic-ray pressure profiles, and CR sound speed ($c_c = \sqrt{p_c/\rho}$) for our $V_g = 200$ simulation (see Table \ref{tab:compare}).   For this plot, because of the very low base gas sound speed, we have normalized the velocity, CR sound speed, and and CR pressure using $V_g$, $V_g$, and $\rho V_g^2$, respectively.   Note the onset of an instability and strong fluctuations at $r \sim 1.07$ (the resolution decreases at $r \simeq 1.14$ due to a change in mesh refinement, which likely is responsible for suppressing the short wavelength fluctuations exterior to that radius).   Despite the large density fluctuations, the mass-loss rate and terminal velocity in the simulation are well-described by our steady state analytic solutions.}
\label{fig:app}
\end{figure}

From equation \ref{eq:amp}, the \citet{Drury1986} instability is most likely to grow when $c_i$ is small and/or $V_g$ is large (gravity is strong), both of which decrease the CR scale-height  (eq. \ref{eq:hc}).  Indeed, we find that our simulation with the smallest value of the gas isothermal sound speed does show evidence of an instability.   In this case (the first row in Table \ref{tab:compare}), we predict $A(r) \simeq 0.5$ near the base of the wind, and a somewhat larger value at the sonic point where $c_c$ is larger.    Figure \ref{fig:app} shows that there is indeed evidence of an instability that sets in at $r \sim 1.07$ in this simulation.   This may be a manifestation of the \citet{Drury1986} instability.   However, the instability in Figure \ref{fig:app} sets in at radii well exterior to the sonic point and even exterior to where the flow speed equals the CR sound speed.  We suspect that these are fluctuations generated by the \citet{Drury1986} instability at small radii and advected out to large radii where they become nonlinear due to conservation of wave action.    Despite the large density fluctuations, however, the wind mass-loss rate and terminal velocity in this simulations are still well-described by the analytic solution in \S \ref{section:analytic}.   We note that the resolution of the simulation in Figure \ref{fig:app} decreases at $r \simeq 1.14$ due to a change in mesh refinement, which likely is responsible for suppressing the short wavelength fluctuations exterior to that radius.

\end{appendix}

\end{CJK*}
\end{document}